\documentclass{aa}  
\usepackage[colorlinks=true,citecolor=blue,linkcolor=blue]{hyperref}

\usepackage{graphicx}
\usepackage{txfonts}

\usepackage{subcaption} 
\usepackage{natbib}
\bibpunct{(}{)}{;}{a}{}{,} 
\usepackage{bm}
\usepackage{upgreek}
\usepackage[dvipsnames]{xcolor}
\usepackage{amsmath}    
\usepackage[super]{nth}
\usepackage{orcidlink}
\usepackage{lineno} 

\newcommand{\nlens}{15}
\newcommand{\newedit}[1]{{#1}} 

\newcommand{\finaledits}[1]{{#1}} 
\newcommand{\finaledit}[1]{{#1}} 

\begin{document} 

   \title{Investigating the relation between environment and internal structure of massive elliptical galaxies using strong lensing}

   \titlerunning{Relation between environment and internal structure of massive ellipticals}
   \authorrunning{Adnan et al.}

   \subtitle{}

   \author{S M Rafee Adnan\inst{\ref{camsust}, \ref{sustphy}, \ref{uofl}}\fnmsep\thanks{These authors contributed equally to this work.}\fnmsep\thanks{Corresponding authors, SMRA: \href{mailto:rafee.adnan.21@gmail.com}{rafee.adnan.21@gmail.com},\newline SHR: \href{mailto:sulymanhossainrobin@gmail.com}{sulymanhossainrobin@gmail.com}.}\orcidlink{0009-0009-4867-098X}
   \and
   Muhammad Jobair Hasan\inst{\ref{sustphy}, \ref{camsust}, \star}\orcidlink{0009-0007-8037-4027}
   \and
   Ahmad Al-Imtiaz\inst{\ref{camsust},\ref{sustphy},\star}\orcidlink{0009-0008-9598-3439}
   \and
   Sulyman H. Robin\inst{\ref{camsust},\ref{sustphy},\star,\star\star}\orcidlink{0009-0005-4283-0154}
   \and
   Fahim R. Shwadhin\inst{\ref{bauet},\star}\orcidlink{0009-0002-4699-8059}
   \and
   Anowar J.~Shajib\inst{\ref{uchicago},\ref{kicp},\ref{cassa}}\fnmsep\thanks{NHFP Einstein Fellow}\orcidlink{0000-0002-5558-888X}
   \and
   Mamun Hossain Nahid\inst{\ref{sustphy}}\orcidlink{0009-0007-3663-8785}
   \and
   Mehedi Hasan Tanver\inst{\ref{camsust},\ref{sustmath}}\orcidlink{0009-0004-5895-6213}
   \and 
   Tanjela Akter\inst{\ref{sustphy}}\orcidlink{0009-0004-7732-2865}
   \and
   Nusrath Jahan\inst{\ref{sustphy},\ref{camsust}}\orcidlink{0009-0004-7069-1780}
   \and
   Zareef Jafar\inst{\ref{iub},\ref{brac}}\orcidlink{0000-0002-4645-6375}
   \and
   Mamunur Rashid\inst{\ref{camsust},\ref{sustphy}}\orcidlink{0009-0006-0184-0628}
   \and
   Anik Biswas\inst{\ref{buet}}\orcidlink{0009-0009-3520-8980}
   \and
   Akbar Ahmed Chowdhury\inst{\ref{cassa},\ref{iubps}}\orcidlink{0009-0007-0989-8012}
   \and
   Jannatul Feardous\inst{\ref{ruet}}\orcidlink{0000-0002-1362-3376}
   \and
   Ajmi Rahaman\inst{\ref{sustmath}}\orcidlink{0009-0003-4208-4089}
   \and
   Masuk Ridwan\inst{\ref{buet}}\orcidlink{0009-0006-3306-2269}
   \and
   Rahul D. Sharma\inst{\ref{sustphy}}\orcidlink{0009-0005-8883-0707}
   \and
   Zannat Chowdhury\inst{\ref{iubcse}}\orcidlink{0000-0003-1436-8749}
   \and
   Mir Sazzat Hossain\inst{\ref{iub},\ref{cassa}}\orcidlink{0000-0001-6999-6879}
   }

    \institute{Copernicus Astronomical Memorial of SUST, Shahjalal University of Science and Technology, Sylhet 3114, Bangladesh \label{camsust}
    \and
    Department of Physics, Shahjalal University of Science and Technology, Sylhet 3114, Bangladesh \label{sustphy}
    \and
    Department of Physics and Astronomy, University of Louisville, Louisville, KY 40208, USA \label{uofl}
    \and 
    Department of Electrical and Electronic Engineering, Bangladesh Army University of Engineering \& Technology, Qadirabad, Natore 6431, Bangladesh\label{bauet}
    \and
    Department of Astronomy \& Astrophysics, University of Chicago, Chicago, IL 60637, USA\label{uchicago}
    \and
    Kavli Institute for Cosmological Physics, University of Chicago, Chicago, IL 60637, USA\label{kicp}
    \and
    Center for Astronomy, Space Science and Astrophysics, Independent University, Bangladesh, Dhaka 1229, Bangladesh \label{cassa}
    \and
    Department of Mathematics, Shahjalal University of Science and Technology, Sylhet 3114, Bangladesh\label{sustmath}
    \and
    Center for Computational \& Data Sciences, Independent University, Bangladesh, Dhaka 1229, Bangladesh\label{iub}
    \and
    Department of Computer Science and Engineering, BRAC University, Dhaka 1212, Bangladesh \label{brac}
    \and
    Department of Electrical and Electronic Engineering, Bangladesh University of Engineering and Technology, Dhaka 1205, Bangladesh\label{buet}
    \and
    Department of Physical Science, Independent University, Bangladesh, Dhaka 1229, Bangladesh\label{iubps}
    \and
    Department of Electrical \& Electronic Engineering, Rajshahi University of Engineering \& Technology, Rajshahi 6204, Bangladesh\label{ruet}
    \and
    Department of Computer Science and Engineering, Independent University, Bangladesh, Dhaka 1229, Bangladesh\label{iubcse}
    }

   \date{Received ; accepted }

  \abstract
  {Strong lensing by massive galaxies probes their mass distribution, thus providing a window to study their internal structure, i.e., the distributions of luminous and dark matter. In this paper, we investigate the relation between the internal structure of massive elliptical galaxies and their environment using a sample of \nlens\ strong lensing systems. We performed lens modeling for them using \textsc{lenstronomy} and constrained the mass and light distributions of the deflector galaxies. We adopt the local galaxy density as a metric for the environment and test our results against several alternative definitions of it. We robustly find that the centroid offset between the mass and light is not correlated with the local galaxy density. This result supports using centroid offsets as a probe of dark matter theories since the environment's impact on it can be treated as negligible. Although we find a moderate to strong correlation between the position angle offset and the standard definition of the local galaxy density, consistent with previous studies, the correlation becomes weaker for alternative definitions of the local galaxy density. This result weakens the support for interpreting the position angle misalignment as having originated from interaction with the environment. Furthermore, we find the `residual shear' magnitude in the lens model to be uncorrelated with the local galaxy density, supporting the interpretation of the residual shear originating, in part, from the inadequacy in modeling the angular structure of the lensing galaxy and not solely from the structures present in the environment or along the line of sight.}


\maketitle


\section{Introduction}

Morphological, structural, star formation, and kinematic properties of galaxies have been found to correlate with their environment \citep{Dressler80, Treu09, Fasano15, Pelliccia19, Marasco23}. Such correlations are considered to emerge from galaxies evolving through interactions, such as mergers, strangulation, harassment, tidal compression \citep[e.g.,][]{Treu03}, with other galaxies in their environment or with the larger-scale halo \citep[e.g.,][for a result from galaxy formation simulation]{Pfeffer23}. In particular, as massive elliptical galaxies are considered to be the end product of hierarchical mergers according to the current paradigm, they are ideal probes to study the connection between their structural properties and the environment they reside in to understand the impact of various interactions on their formation and evolution.

Strong gravitational lensing, the phenomenon of multiple images of a background source forming due to the gravitational bending of the light path, provides a powerful tool to probe the internal structure of the massive lensing objects (e.g., galaxies, groups, or clusters) in the foreground. Whereas the observed light distribution traces only the luminous matter distribution, strong lensing probes the total matter distribution. As a result, strong lenses make excellent candidates for studying the structural properties, especially alignment or misalignment between the mass and light profiles, arising from the alignment or misalignment between the dark matter and the baryons, and its connection with the environment.

Numerous previous studies used samples of strong lensing galaxies to constrain their structural properties. These studies mostly found tight alignment between the position angles (PAs) of the mass and light distributions (within $\sim$10$\degr$) \citep{Keeton98b, Kochanek02, Treu09, Gavazzi12, Sluse12, Bruderer16, Shajib19, Shajib21}. The cases with higher misalignments also coincided with a larger than typical `external' shear magnitude, although the vice versa is not always true. `External' shear in the literature has been commonly attributed to physical origin from the nearby galaxies or structures along the line of sight. Hence, the presence of a large `external' shear with a large misalignment could be interpreted as systems in a crowded environment that are not yet dynamically relaxed, as these can have stellar orbits misaligned with the underlying dark matter distribution, given that simulations have found highly misaligned orbits in isolated systems to be unstable and rare \citep{Heiligman79, Martinet88, Adams07, Debattista15}. \citet{Treu03} indeed found a significant ($\sim$3$\sigma$) correlation between the local galaxy density \citep{Dressler80, Cooper05} and the PA misalignment. However, \citet{Etherington23} brought into question the interpretability of the `external' shear to have physical origin, as they illustrate that a large `external' shear magnitude in the best-fit lens model can arise from the inadequacy of the mass model for lensing galaxy in capturing all of its angular complexity \citep[for example, boxy- or discy-ness, ellipticity gradient, isophotal twists;][]{VandeVyvere22, VandeVyvere22b}. For that reason, \citet{Shajib24} recommend residual shear as a more appropriate nomenclature for the additional shear field commonly included in lens models. However, the lack of physical interpretability of the shear magnitude constrained by lens modeling makes the explanation of high misalignment stemming from interaction with a crowded environment less secure. Therefore, it is necessary to investigate the connection between mass and light misalignment with the environment through direct measures, such as local galaxy density, which we set as the goal of this paper.

Additionally, the offset between the mass and light centroid can also provide valuable insights into the nature of the dark matter. Cold dark matter (CDM) simulations predict that the offset between the mass and light centroids should be small \citep[$\lesssim$600 pc for 95\% of the galaxies;][]{Schaller15}. On the contrary, the self-interacting dark matter (SIDM) theory predicts a larger offset between the mass and light centroids \citep{Harvey14, Kahlhoefer14, Robertson17}. However, if such offsets correlate with the environment, it could indicate that the offset is not due to the dark matter self-interaction but due to the galaxy's interaction with the environment. For that reason, we also aim in this paper to investigate the connection between the centroid offset between the mass and light with the environment. If a lack of correlation between the centroid offset and the environment is observed, then it would help establish such offsets as a robustly testable prediction of the SIDM theory.

In this paper, we model a sample of \nlens\ strong lensing galaxies from Hubble Space Telescope (HST) imaging to constrain their mass and light misalignments (or the lack thereof) and investigate their correlation with the local galaxy density. Our sample is advantageous for this particular science question for including systems with larger Einstein radii than typically found in previous samples. The Einstein radius, $\theta_{\rm E}$ range of our sample is \newedit{0\farcs86--3\farcs32}, whereas the one for the Sloan Lens ACS Survey \citep[SLACS;][]{Bolton06} sample is \newedit{0\farcs69--1\farcs78} \citep{Auger09}, and that for the Strong Lensing Legacy Survey \citep[SL2S;][]{Gavazzi12} with available HST imaging is \newedit{0\farcs71--2\farcs58} \citep{Tan24, Sheu24}. The systems with larger Einstein radii tend to be groups with multiple members existing within the Einstein radius, which makes this sample better for probing the structural properties of galaxies in locally denser environments. The lens models we present in this paper are the first in the literature for these systems, which will enable future studies of their dark matter halos in combination with follow-up measurements such as the stellar kinematics \citep[e.g.,][]{Tran22, Tan24, Sheu24}.

This paper is organized as follows. In Section \ref{sec:data}, we describe the HST data we modeled. Then, in Section \ref{sec:lens_modeling}, we describe our lens modeling method and our lens sample. We present our results in Section \ref{sec:result} and discuss them in Section \ref{sec:discussion}. Finally, we summarize and conclude the paper in Section \ref{sec:summary}. Throughout the paper, we adopt a flat $\Lambda$CDM cosmology as the fiducial cosmology with $H_0= 70$ km s$^{-1}$ Mpc$^{-1}$ and $\Omega_{\rm m} = 0.3$.

\section{HST imaging data} \label{sec:data}

\begin{figure*}
	\includegraphics[width=\textwidth]{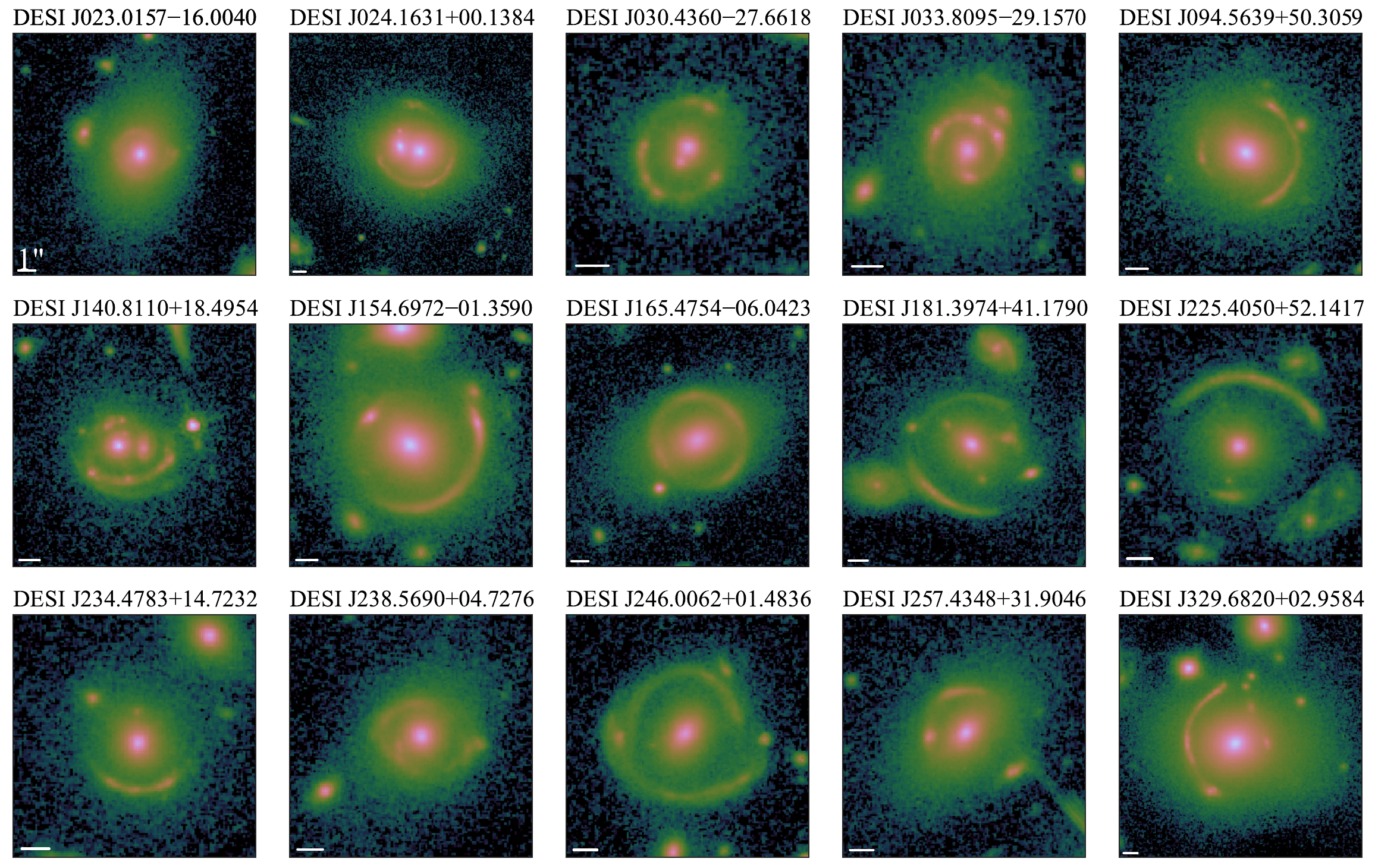}
	\caption{\label{fig:montage}
Montage of \nlens\ strong lensing systems modeled in this paper. These systems were identified as lens candidates using a deep neural network from the DESI Legacy Imaging Surveys data. Their lensing nature was then confirmed through high-resolution HST imaging displayed here. Each image was obtained with the WFC3 IR channel in the F140W filter. Up is aligned with the \newedit{north}, and the left-hand direction is aligned with the east in these images. The white bar in each panel indicates a scale of 1\arcsec.}
\end{figure*}

Our sample comprises \nlens\ strong lensing systems (illustrated in Fig. \ref{fig:montage}). These systems were discovered as lens candidates in the Dark Energy Spectroscopic Instrument (DESI) Legacy Imaging Surveys, Data Release 7 \citep{2019AJ....157..168DNew}, using a deep neural network \citep{2020ApJ...894...78H}. They were then confirmed using the follow-up HST imaging from the program SNAP-15867 (PI: Huang, \citealt{bib3:Huang25}).

The HST imaging was obtained using the Wide Field Camera 3 (WFC3) in the infrared (IR) channel using the F140W filter. The total exposure time for each image was 1197.7 \newedit{seconds}.

We produced science-quality reduced images by combining multiple exposures using the \textsc{astrodrizzle} software package \citep{Avila15}.\footnote{The data reduction procedure followed the \textsc{jupyter} notebooks from this GitHub repository: \url{https://github.com/ajshajib/hst-lens}.} The drizzling process with \textsc{astrodrizzle} removed cosmic rays. We chose 0\farcs08 as the pixel scale of the final image after drizzling.
The final images were rotated through the drizzling process to make the north and east directions align with the vertical and horizontal directions, respectively. We used the \textsc{Python} package \textsc{PhotUtils} to estimate the mean background level and subtracted it from each image. The total noise per pixel was estimated by summing the background root-mean-square (RMS) level in quadrature with the Poisson noise corresponding to the background-subtracted-flux in each pixel.

\section{Lens modeling} \label{sec:lens_modeling}
We used the publicly available software package \textsc{lenstronomy}\footnote{\url{https://github.com/lenstronomy}} \citep{Birrer15, bib3:Birrer18} to model all the lenses in our sample. The lens modeling program \textsc{lenstronomy} has proven to be effective, as demonstrated in the Time-Delay Lens Modeling Challenge, where various teams successfully used the software to recover the lens parameters \citep{Ding21}. Additionally, \textsc{lenstronomy} has been applied in numerous areas related to strong gravitational lensing, including time-delay cosmography \citep{Birrer19, Shajib20, Shajib22b} and the study of dark-matter substructures \citep{Gilman20}.

In section \ref{ssec:lens_model_ingredients}, we discuss all the model components used to describe the mass and light distributions of the deflector and source galaxies. In section \ref{modeling_procedure}, we describe our procedure for lens modeling and obtaining the posterior probability distribution function (PDF) of each model parameter. Finally, in section \ref{ssec:specific_descriptions}, we describe specific characteristics of each lens system that informed the choices we made in setting up the lens model.

\subsection{Lens model ingredients}
\label{ssec:lens_model_ingredients}

We adopted a baseline model for the lens mass and light distribution that works well for the majority of lensing systems in our sample of different sizes and shapes. This baseline model is then adjusted to adapt to the specific complexity of a given system when necessary. Our goal for the baseline model is to fit the data nicely while being flexible enough for a wide range of lens systems.

\subsubsection{Mass profile} \label{sec:mass_profile}

\paragraph{EPL:}

The convergence for the elliptical power law \citep[EPL;][]{Tessore15} mass profile is given by

 \begin{equation}
\kappa\left(x, y\right)=\frac{3-\gamma}{2}\left(\frac{\theta_{\mathrm{E}}}{\sqrt{q x^2+y^2 / q}}\right)^{\gamma-1}.
\end{equation}    

Here, \(\theta_{\mathrm{E}}\) is the Einstein radius. The parameter \(q\) denotes the axis ratio of the ellipse, representing the flattening of the projected mass distribution. In this formulation, the coordinate system is assumed to be aligned with the major and minor axes, where the major axis can have a PA with respect to the on-sky coordinate frame.

\paragraph{SIE:} In situations where a nearby satellite galaxy has a notable impact on the lensing effects experienced by source galaxies, we adopt a singular isothermal ellipsoid (SIE) profile to characterize its mass distribution, which is a special case of the EPL mass profile with \(\gamma=2\).

\paragraph{Residual Shear:} We adopt the residual shear field to account for any additional shear that is not captured by the EPL or SIE profiles.
The residual shear field is parametrized with two parameters: the shear magnitude $\gamma_{\rm shear}$ and angle $\phi_{\rm shear}$.

\paragraph{Flexion:} The second-order lensing effect can be expressed with four flexion terms \citep{Schneider08}, which are the third derivatives of the lensing potential. The flexion is responsible for introducing a curvature and other anisotropic distortions
in the images.

\subsubsection{Light profiles} \label{sec:light_profile}

\paragraph{Elliptical S\'ersic:}
We chose the elliptical S\'ersic function \citep{Sersic68} to model the deflector light profile. The S\'ersic profile is parameterized as

\begin{equation}
I\left(x, y\right)=I_{\mathrm{e}}\, \exp \left[-k\left\{\left(\frac{\sqrt{x^2+y^2 / q_{\mathrm{L}}^2}}{R_{\mathrm{eff}}}\right)^{1 / n_{\text {S\'ersic }}}-1\right\}\right]
.
\end{equation}

where $R_{\rm eff}$ is the effective radius, $I_{\rm e}$ is the surface brightness at $R_{\rm eff}$, $q_{\rm L}$ is axis ratio, $n_{\text{S\'ersic}}$ is the S\'ersic index, and $k$ is a normalizing constant so that $R_{\rm eff}$ becomes the half-light radius.

\paragraph{Shapelets:}
Along with the elliptical S\'ersic profile, we used a basis set of shapelets (i.e., 2D Gauss--Hermite polynomials) to describe the light distribution of the source galaxy. The maximum polynomial order $n_\text{max}$, determines the number of shapelet components $N_\text{shapelets}$, with $N_\text{shapelets} = (n_\text{max} + 1)(n_\text{max}+2)/2$. Initially, we set $n_\text{max} = 6$ as the default for all systems. However, for systems with more complex structures on the lensed arcs, we increased the order $n_\text{max}$ through trial and error to enhance the model’s goodness of fit. The specific $n_\text{max}$ values chosen for each system are mentioned in Table \ref{table:lens_profiles}.

\subsection{Modeling procedure}

\label{modeling_procedure}

In this subsection, we describe our modeling procedure: setup of the image cutout and point spread function (PSF) in Section \ref{sec:PSF}, masking and initial optimization in Section \ref{sec:optimization}, and running the Markov Chain Monte Carlo (MCMC) method in Section \ref{sec:mcmc}.

\subsubsection{Setting up image cutouts and PSF}\label{sec:PSF}

We chose an image cutout encompassing the lens and its immediate surroundings from the entire HST image. We used the same pixelated PSF for all the lens systems, which was produced using the \textsc{TinyTim} software program \citep{Krist11}. A circular or elliptical mask is created depending on the system's shape to exclusively encompass the deflector light distribution and the associated arcs.
In cases where there exists a nearby galaxy or star, these are deliberately masked out unless a specific decision is made to model the light profile of a satellite, for example, for DESI J030.4360$-$27.6618 as described in Section \ref{model:2739}.

\subsubsection{Initial model optimization}\label{sec:optimization}

After setting up the image cutouts and masks as described above, we ran the Particle Swarm Optimization (PSO) procedure for the baseline model setup. Then, we iteratively fine-tuned (i.e., through a trial-and-error process) our model setup as required by the complexity of the lens system until we achieve the desired goodness of fit. PSO is suited for this sort of iterative fine-tuning phase as it is computationally much cheaper than sampling methods such as the MCMC.
 
The considered sample of the gravitational lenses contains lenses with various levels of complexity. Some of these were quite simple, and the selected baseline model was sufficient for their modeling. In contrast, others needed increasingly complex combinations of lens-light, lens-mass, and source-light profiles. 

The most common case for adding complexity was adding a second S\'ersic profile to the existing one describing the deflector's light to mitigate large residuals in the center. For these cases, we joined the centroid of the second S\'ersic profiles with that of the first. We also joined the ellipticity parameters between the two S\'ersic profiles, except for the systems DESI J238.5690$+$04.7276 and DESI J329.6820$+$02.9584, as they required more flexibility for this parameter to give a good model fit.
 
In some systems, our initial model provided a mass distribution with unusually high ellipticity compared to its corresponding light distribution. However, a small difference is expected between them, with the mass being generally rounder than the light \citep[e.g.,][]{bib3:Schmidt23, Sheu24}. To reduce this discrepancy, we imposed a prior condition to incentivize the mass distribution to be rounder than the light. While this prior resulted in a more plausible model regarding the mass ellipticity, other major parameters were not significantly altered.

Another case of multiple occurrences was for the satellites located near the central lensing galaxies. Masking out these satellites may not be productive as it would result in the loss of valuable lensing information, and that would also disallow us to account for the lensing perturbations caused by the satellites. In these cases, satellites were modeled using the SIE mass and the elliptical S\'ersic light profiles. We joined the centroids of the mass and light profiles for these satellites. In some cases, the lensing galaxies were in environments crowded with satellite galaxies or companions that would require too many additional mass profiles or too much complexity to explicitly model. However, to account for the higher-order lensing effects of these objects, we added a flexion field.

The baseline source-light model consisted of a basis set of shapelets and one elliptical S\'ersic profile. To account for extra source components, such as additional blobs or arcs distinct from the primary set of arcs or rings at the Einstein radius, additional source-light profiles (e.g., another set of shapelets) were introduced. Another instance of implausible lens models during the trial-and-error phase was when the modeled source galaxy resembled a scaled-down version of the lensed arcs or rings. Given that an arc- or ring-shaped source galaxy is highly improbable, this suggests that those best-fit lens models assume an unreasonable shape and size (larger than the truth) for the source galaxy, indicative of left-over unaccounted deflections (i.e., lensing power) leading to the predicted Einstein radius being smaller than the truth. To guide the modeling in the right direction, we constrained the sources to much smaller sizes.

We considered the following general criteria for accepting a lens model to have reached an optimal setup after fine-tuning: (i) the model can reconstruct prominent lensing features observed in the HST image, (ii) model residuals are better than a set threshold \newedit{(i.e., $\chi_{\rm red}^2 < 1.2$)} ensuring an acceptable level of source reconstruction, (iii) the reconstructed source does not bear discernible similarities to the lensed arc (as discussed at the end of the previous paragraph), (iv) the logarithmic slope $\gamma$ is within the range \newedit{$1.4 < \gamma < 2.8$ \citep[following][]{Tan24}}, and (v) the model is not over-fitting (i. e., the reconstructed source must not \newedit{contribute largely to the total light observed at the deflector center}). We proceeded with the model setups that fulfill these criteria for further optimization with the MCMC method.

\subsubsection{MCMC sampling}\label{sec:mcmc}

After completing the PSO with a viable model, the subsequent step involved executing the MCMC sampling procedure. Commencing from the best-fit results obtained through the PSO, we continued this sampling until a satisfactory level of convergence of the parameter values was achieved. This specific initialization of the parameter values facilitates a more rapid convergence of the MCMC chain. We examined the trace of the MCMC walkers and considered a stable distribution of the walker positions for at least 1000 steps for all the non-linear parameters as the criteria for convergence. Finally, a comprehensive reassessment of the model's goodness of fit was conducted by inspecting the model residuals and the parameter values, a procedure similar to the approach employed during our fine-tuning procedure in Section \ref{sec:optimization}.

\begin{figure*}
	\centering
	\includegraphics[width=1.4\textwidth]{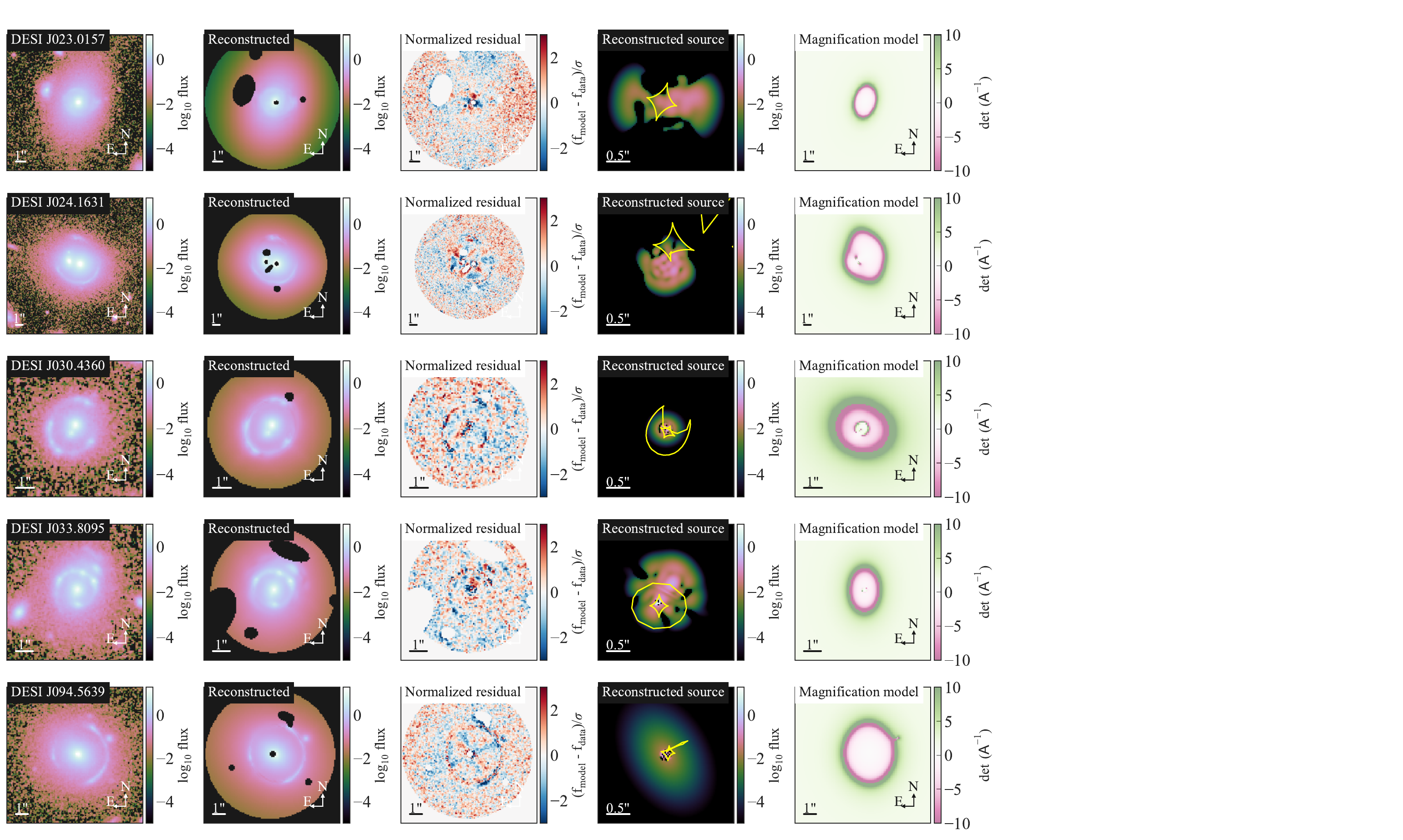}
	\caption{\label{fig:lens_models_0}
Illustration of lens models for the first five out of 15 systems in our sample. First column: Image cutouts of the lensing systems. Second column: Reconstructed images based on the optimized lens model. The dark regions are masked-out pixels. Third column: Residual maps normalized by the noise levels in each pixel. Fourth column: Reconstructed flux distribution of the unlensed source galaxy. The yellow lines show the caustics. Fifth column: The magnification model, computed as the inverse of the determinant of the Jacobian matrix, reveals spatial variations in magnification due to the gravitational lensing effect. For all the lens systems, major lensing features, such as the prominent arcs, are successfully reproduced. Whereas some residuals above the noise levels are present in some cases, they would require a larger complexity in the source description (e.g., increasing the shapelet order $n_{\rm max}$) and thus significantly increase the computational time. From tests done on a few of the complex systems (e.g., DESI J246.0062$+$01.4836), we find that increasing the complexity further does not substantially alter the lens model parameters of interest. Furthermore, the goodness-of-fit of these models is typical and sufficient for the science goals and similar ones in the literature \citep[cf.,][]{Shajib21}.
	}
\end{figure*}

\subsection{Specific descriptions of the lens systems}
\label{ssec:specific_descriptions}

In this subsection, we describe the specific morphology and characteristics pertaining to each \newedit{strong lensing system} and state the lens model settings required to account for them. \newedit{We also summarize the specific model settings for each system in the appendix (Table \ref{table:lens_profiles}).}

\subsubsection{DESI J023.0157$-$16.0040}

\label{model:0132}
In this system, a faint arc surrounds the central deflector galaxy. The central deflector is accompanied by a large neighboring galaxy to the east and a few other smaller galaxies. We assumed the lensing effect of these neighboring galaxies to be negligible, given their relatively small size and sizable distance from the arcs, and masked over their light. We also noticed some concentrated residuals in the center of the main deflector galaxy's light, which the double S\'ersic profiles were not adequate to describe. Therefore, we put a small circular mask at the center of the main deflector galaxy.

\subsubsection{DESI J024.1631$+$00.1384}\label{model:0008}

This system has two lensing galaxies of comparable sizes within the Einstein radius. Another smaller object also exists within the Einstein radius, which is potentially a third member within this group. We modeled the larger galaxy on the west as the main deflector. We also explicitly modeled the mass profile of the large companion with the SIE model. However, we assumed the lensing effect of the potential third group member is negligible, given its relatively smaller size and its distance from the arcs. Therefore, we masked over its light. \newedit{We imposed a prior constraint on the companion's Einstein radius, requiring it to scale with its mass under the assumption that the mass-to-light ratio is identical for both the main deflector and the companion. This constraint is expressed as $\theta_{\rm E, Cen}/\theta_{\rm E, Sat} = \sqrt{{\rm flux_{Cen}}/{\rm flux_{Sat}}}$, where, the subscript `Cen' corresponds to the central larger deflector and the subscript `Sat' corresponds to the companion}. This relation follows from the assumption that the total mass ($M$) of each deflector scales proportionally with its luminosity and the fact that $\theta_{\rm E} \propto \sqrt{M}$. We also noticed some prominent residual in between these two galaxies that the double S\'ersic profiles were not adequate to describe. This non-smooth structure may potentially arise if these two galaxies are interacting or merging. We masked this non-smooth structure in the light distribution (see Fig. \ref{fig:lens_models_0}).

\subsubsection{DESI J030.4360$-$27.6618} \label{model:2739}

This system features a lensing arc closely resembling an ellipse, with two central galaxies within the Einstein radius. We modeled the mass distribution of both galaxies. We treated the larger galaxy as the primary deflector and used the SIE mass profile for the secondary deflector \newedit{galaxy as a satellite}. There is a small, faint object near the lensing arc in the north. Although it is unclear whether it is part of the source galaxy, we masked it to simplify the model (see Fig. \ref{fig:lens_models_0}). We also imposed a prior condition on the satellite galaxy's Einstein radius similar to the system DESI J024.1631$+$00.1384, described above in Section \ref{model:0008}.

\subsubsection{DESI J033.8095$-$29.1570}

In this system, there is only one lensing galaxy inside the Einstein radius. Additionally, we observe another arc-like object northwest of the main arc. In our model, we recreated both arcs by assuming two source components, which were modeled with two separate sets of shapelets. Although there is a possibility of the outer arc being an edge-on disk galaxy located along the line of sight, adding a second source improved the residual on the counter-arcs in the southeast as well. Thus, we concluded that our model setup with the two source components is appropriate. There is also a faint arc-like object near the northwest of the lensing arc, which we masked out so as not to require further complexities in the lens model.

\subsubsection{DESI J094.5639$+$50.3059}

This system is accompanied by a few smaller galaxies, some situated close to the arc. We assumed the lensing effect for most of these neighboring galaxies to be negligible, given their relatively smaller size and distance from the arcs, and masked their light accordingly. However, through initial trial and error to fine-tune the model, we find that the relatively bright galaxy to the northwest needs to be explicitly modeled as a satellite galaxy. \newedit{We imposed a prior condition on the satellite galaxy's Einstein radius similar to the system DESI J024.1631$+$00.1384, as described in Section \ref{model:0008}.} Similar to \ref{model:0008}, we also masked the center of the main deflector galaxy to account for the modeling insufficiency in the case of using two superimposed S\'ersic profiles.

\subsubsection{DESI J140.8110$+$18.4954}

This system has two central deflectors. We considered the eastern one as the main deflector and the western deflector as a satellite galaxy. To keep the complexity of the source description at a computationally feasible level, we masked the fainter extra arc beneath the more prominent one. We also masked out a region surrounding the satellite lens galaxy, as the residuals in that region would not improve with further complexity provided in the satellite's light model. Additionally, we masked out a star in the northwest and a small blob in the northeast that is potentially a line-of-sight galaxy with negligible impact. 

\subsubsection{DESI J154.6972$-$01.3590}

This system has two blobs around the northeast and the southwest corners. The shapes and locations of these blobs make them very unlikely to be lensed features. Therefore, we considered them as line-of-sight structures and masked them out.

\subsubsection{DESI J165.4754$-$06.0423}

Similar to some other systems, we used double S\'ersic profiles to model the light profile of the central galaxy. However, the double S\'ersic profile is not sufficient to model the center of the deflector. Therefore, we masked out a small circular region at the center of the main deflector.

\subsubsection{DESI J181.3974$+$41.1790}

In this system, a prominent central galaxy is accompanied by multiple smaller galaxies positioned within its Einstein radius. We find the nearby small galaxies within the Einstein radius to have negligible lensing effects. Therefore, we masked out their lights and excluded them from the mass model. We added a flexion field, which may have originated from the large spiral galaxies located just outside the arcs. This was necessary to accurately reproduce the distortions in the arcs, as the central deflector alone -- even after including the mass profiles of smaller galaxies -- was insufficient to reproduce their morphology to the required level.

\subsubsection{DESI J225.4050$+$52.1417}

This system has a little faint blob just south of the central deflector within the Einstein radius. We assumed this blob to have an insignificant contribution to the lensing effects due to its small size and consequently masked it. Although there are multiple galaxies outside the arcs, we masked out all of them as the combination of the central deflector's mass and residual shear was sufficient to reproduce the lensed features.

\subsubsection{DESI J234.4783$+$14.7232}

There are multiple galaxies near the line of sight in this system, with the smaller ones close to the arcs and the larger ones further away. Since such a combination of distance and mass for the nearby line-of-sight galaxies serendipitously makes their individual lensing contribution negligible, we masked all of them and ignored their mass profiles in our lens model.

\subsubsection{DESI J238.5690$+$04.7276}

In this system, two blobs are observed in the northeast and southwest regions, which are relatively far away from the lensed arcs. We masked them and excluded their mass from our lens model.

\subsubsection{DESI J246.0062$+$01.4836}
 
In our model for this system, we considered the blob at the west near the arc to be a satellite galaxy. A faint blob near the central deflector (in the southeast) was masked because we assumed its effect on the lensing system to be minimal. Furthermore, we masked out some parts of the southwestern portion of the arc, as that region exhibited strong residuals that would require significantly higher complexity in the source description to resolve. However, we find from a test that increasing the source complexity does not significantly alter the lens model parameters.

\subsubsection{DESI J257.4348$+$31.9046}

We masked out a small blob in the west of this system. Additionally, there are some elongated features in the southwest, identified as diffraction spikes from a nearby star that is situated outside the cutout region. We masked out these spikes.

\subsubsection{DESI J329.6820$+$02.9584}

In this system, there are several neighboring galaxies of various sizes in the north. We assumed their contribution to the lensing effects to be negligible due to their combination of size and distance from the lensing arc. This assumption is justified by successfully reproducing the lensed arcs with our baseline model setup. \newedit{In addition, we masked the center of the main deflector galaxy to address the inadequacy in modeling using two superimposed S\'ersic profiles (similar to \ref{model:0008}).}

\section{Result} \label{sec:result}

In this section, we present the best-fit values of the lens model parameters for our sample of modeled systems in Section \ref{sec:lens_params} and our estimates of the local galaxy densities around the lenses in Section \ref{sec:local_densities}. We then present our results on the alignment between mass and light in Section \ref{sec:alignment_results}. Finally, in Section \ref{sec:lens_model_params}, we discuss the correlation between the local galaxy density and lens
model parameters.

\subsection{Lens model parameters} \label{sec:lens_params}

We present the point estimates of the lens model parameters with their $1\sigma$ uncertainties in Table \ref{table:lens_params}. The 1$\sigma$ uncertainty levels were extracted from the \nth{16} and \nth{84} percentiles of the values sampled from the posterior PDF using the MCMC method. The optimized lens models are illustrated in Figures \ref{fig:lens_models_0}, \ref{fig:lens_models_1}, and \ref{fig:lens_models_2}.

\renewcommand{\arraystretch}{1.3}
 \begin{table*}
 \centering
  \begingroup
	\setlength{\tabcolsep}{3pt} 
	\renewcommand{\arraystretch}{1.3} 
 \caption{
 Point estimates of the lens model parameters. \label{table:lens_params}
}

\centering
\begin{tabular}{lccccccccc}
\hline
     System &  $\theta_{\rm E}$ &    $\gamma$ &    $q_\text{m}$ &     $\phi_\text{m}$ &  $\gamma_\text{shear}$  &  $\phi_\text{shear}$  &
     $R_{\rm eff} $ & 
     $q_\text{L}$ & 
     $\phi_\text{L}$
     \\
     & (\arcsec) & & &(\degr)   & & (\degr) & (\arcsec) & & (\degr) \\
\hline
J023.0157$-$16.0040 &         $1.403_{-0.006}^{+0.005}$ &         \finaledits{--} &         $0.676_{-0.020}^{+0.023}$ &         $-78.5_{-2.6}^{+2.5}$ &         $0.026_{-0.008}^{+0.009}$ &         $52.3_{-8.5}^{+8.6}$ &         $1.07 \pm 0.02$ &         $0.713_{-0.002}^{+0.002}$ &         $-74.4_{-0.2}^{+0.2}$ \\ 
J024.1631$+$00.1384 & $2.711_{-0.009}^{+0.008}$ &$2.01_{-0.04}^{+0.04}$ &         $0.538_{-0.004}^{+0.005}$ &         $70.0_{-0.3}^{+0.3}$ &         $0.134_{-0.003}^{+0.003}$ &         $62.1_{-0.5}^{+0.5}$ &         $0.98 \pm 0.02$ &         $0.900_{-0.001}^{+0.001}$ &         $10.5_{-0.3}^{+0.3}$ \\ 
J030.4360$-$27.6618 &         $0.867_{-0.016}^{+0.013}$ &         $1.62_{-0.03}^{+0.03}$ &         $0.948_{-0.010}^{+0.008}$ &         $-1.5_{-6.7}^{+5.6}$ &         $0.077_{-0.003}^{+0.004}$ &         $-48.9_{-1.0}^{+1.1}$ &         $0.57 \pm 0.01$ &         $0.972_{-0.007}^{+0.007}$ &         $-46.3_{-11.9}^{+10.2}$ \\ 
J033.8095$-$29.1570 &         $1.055_{-0.001}^{+0.002}$ &         \finaledit{$1.80_{-0.04}^{+0.04}$} &         $0.717_{-0.005}^{+0.008}$ &         $-88.8_{-0.4}^{+0.5}$ &         $0.011_{-0.003}^{+0.002}$ &         $-81.8_{-4.4}^{+5.6}$ &         $0.92 \pm 0.02$ &         $0.794_{-0.004}^{+0.004}$ &         $-78.9_{-0.7}^{+0.6}$ \\ 
\finaledit{J094.5639$+$50.3059} &         $2.265_{-0.001}^{+0.001}$ &         \finaledits{--} &         $0.726_{-0.002}^{+0.003}$ &         $83.3_{-0.3}^{+0.3}$ &         $0.074_{-0.001}^{+0.001}$ &         $79.1_{-0.4}^{+0.5}$ &         $0.82 \pm 0.02$ &         $0.792_{-0.002}^{+0.002}$ &         $31.4_{-0.3}^{+0.4}$ \\ 
\finaledit{J140.8110$+$18.4954} &         $1.408_{-0.010}^{+0.011}$ &         \finaledit{$2.22_{-0.04}^{+0.04}$} &         $0.956_{-0.013}^{+0.011}$ &         $44.5_{-15.4}^{+16.0}$ &         $0.126_{-0.004}^{+0.005}$ &         $-10.7_{-0.7}^{+0.9}$ &         $0.54 \pm 0.01$ &         $0.986_{-0.004}^{+0.005}$ &         $-21.6_{-11.6}^{+11.9}$ \\ 
\finaledit{J154.6972$-$01.3590} &         $2.915_{-0.002}^{+0.002}$ &         \finaledit{$1.49_{-0.03}^{+0.03}$} &         $0.727_{-0.005}^{+0.005}$ &         $45.5_{-0.3}^{+0.3}$ &         $0.079_{-0.002}^{+0.002}$ &         $45.7_{-0.5}^{+0.6}$ &         $1.23 \pm 0.02$ &         $0.815_{-0.002}^{+0.001}$ &         $37.3_{-0.2}^{+0.2}$ \\ 
\finaledit{J165.4754$-$06.0423} &         $2.618_{-0.002}^{+0.002}$ &         \finaledits{--} &         $0.675_{-0.008}^{+0.006}$ &         $-22.4_{-0.7}^{+0.5}$ &         $0.038_{-0.003}^{+0.003}$ &         $-5.7_{-1.5}^{+1.4}$ &         $1.33 \pm 0.03$ &         $0.631_{-0.002}^{+0.001}$ &         $-23.6_{-0.2}^{+0.2}$ \\ 
\finaledit{J181.3974$+$41.1790} &         $2.815_{-0.007}^{+0.007}$ &         \finaledit{$2.05_{-0.04}^{+0.04}$} &         $0.631_{-0.009}^{+0.008}$ &         $27.3_{-0.7}^{+0.7}$ &         $0.096_{-0.005}^{+0.004}$ &         $39.7_{-1.0}^{+0.8}$ &         $0.89 \pm 0.02$ &         $0.793_{-0.002}^{+0.003}$ &         $39.0_{-0.4}^{+0.5}$ \\ 
\finaledit{J225.4050$+$52.1417} &         $2.603_{-0.004}^{+0.004}$ &         \finaledit{$2.08_{-0.04}^{+0.04}$} &         $0.830_{-0.022}^{+0.026}$ &         $-4.0_{-2.7}^{+2.1}$ &         $0.147_{-0.009}^{+0.008}$ &         $5.1_{-0.6}^{+0.6}$ &         $0.72 \pm 0.01$ &         $0.953_{-0.005}^{+0.005}$ &         $26.9_{-3.3}^{+3.0}$ \\ 
\finaledit{J234.4783$+$14.7232} &         $1.552_{-0.016}^{+0.006}$ &         \finaledit{$2.06_{-0.08}^{+0.07}$} &         $0.656_{-0.017}^{+0.025}$ &         $90.0_{-0.7}^{+1.3}$ &         $0.025_{-0.004}^{+0.013}$ &         $-42.9_{-17.6}^{+16.2}$ &         $0.71 \pm 0.01$ &         $0.880_{-0.004}^{+0.004}$ &         $76.6_{-0.9}^{+0.9}$ \\ 
\finaledit{J238.5690$+$04.7276} &         $1.502_{-0.005}^{+0.004}$ &         \finaledit{$1.85_{-0.04}^{+0.04}$} &         $0.800_{-0.011}^{+0.013}$ &         $-55.3_{-2.3}^{+2.2}$ &         $0.106_{-0.006}^{+0.005}$ &         $6.8_{-1.3}^{+1.5}$ &         $0.78 \pm 0.02$ &         $0.919_{-0.002}^{+0.002}$ &         $-49.4_{-0.7}^{+0.8}$ \\ 
\finaledit{J246.0062$+$01.4836} &         $2.683_{-0.004}^{+0.004}$ &         \finaledit{$1.71_{-0.05}^{+0.05}$} &         $0.828_{-0.007}^{+0.008}$ &         $-50.2_{-0.8}^{+0.8}$ &         $0.041_{-0.004}^{+0.003}$ &         $-42.8_{-1.3}^{+1.3}$ &         $0.79 \pm 0.02$ &         $0.698_{-0.003}^{+0.002}$ &         $-56.4_{-0.3}^{+0.4}$ \\ 
\finaledit{J257.4348$+$31.9046} &         $2.029_{-0.004}^{+0.004}$ &         \finaledit{$1.74_{-0.03}^{+0.03}$} &         $0.678_{-0.012}^{+0.011}$ &         $-36.3_{-0.5}^{+0.6}$ &         $0.040_{-0.004}^{+0.004}$ &         $-18.2_{-2.5}^{+2.8}$ &         $0.89 \pm 0.02$ &         $0.619_{-0.002}^{+0.002}$ &         $-46.7_{-0.2}^{+0.2}$ \\ 
\finaledit{J329.6820$+$02.9584} &         $3.319_{-0.004}^{+0.004}$ &         \finaledit{$1.94_{-0.04}^{+0.04}$} &         $0.763_{-0.004}^{+0.004}$ &         $-18.3_{-0.6}^{+0.7}$ &         $0.112_{-0.005}^{+0.005}$ &         $-75.1_{-0.8}^{+1.0}$ &         $1.22 \pm 0.02$ &         $0.905_{-0.000}^{+0.001}$ &         $-8.1_{-0.1}^{+0.2}$ \\ 
\hline

\end{tabular}
\tablefoot{\finaledit{The tabulated parameters are the Einstein radius $\theta_{\rm E}$, logarithmic slope of the mass profile $\gamma$, the mass axis ratio $q_{\rm m}$, the major axis position angle for mass $\phi_{\rm m}$, the residual shear magnitude $\gamma_{\rm shear}$, the residual shear angle $\phi_{\rm shear}$, the effective radius (i.e., half-light radius) of the light profile $R_{\rm eff}$, the light axis ratio $q_{\rm L}$, and the major axis position angle for light $\phi_{\rm L}$. The angles $\phi_{\rm m}$, $\phi_{\rm shear}$, and $\phi_{\rm L}$ are defined as north of east. The point estimates are obtained from the medians of the 1D marginalized posteriors, and the 1$\sigma$ uncertainties are obtained from the \nth{16} and \nth{84} percentiles. \finaledit{The reported $R_{\rm eff}$ for the lenses with a double S\'ersic light profile was obtained by numerically solving for the half-light radius, and the effective $q_{\rm L}$ for them was computed from the quadrupole moments of the double S\'ersic profile. The statistical uncertainty levels are typical for lens models based on HST imaging \citep[cf.][]{Tan24}. \finaledit{However, for $R_{\rm eff}$ and $\gamma$, we added 2\% systematic uncertainty levels in quadrature following \citet{Shajib19, Shajib21}. The systems with missing $\gamma$ measurements have them fixed at $\gamma=2$ in their lens models.}}}}
\endgroup
\end{table*}

\renewcommand{\arraystretch}{1.3}
 \begin{table*}

 \caption{\finaledit{Photometric redshifts of the deflector galaxies ($z_{\rm d}$), spectroscopic redshifts of the source galaxies ($z_{\rm s}$)}, local galaxy densities, and centroid offset and position angle misalignment \finaledit{($\Delta \phi$)} between the mass and light. 
\label{table:photometric_params}
}

\begin{centering}
 \begingroup
	\setlength{\tabcolsep}{4pt} 
	\renewcommand{\arraystretch}{1.3} 
    
\begin{tabular}{lccccccccc}
\hline
     System & 
     \finaledit{$z_{\rm d}$} & 
     \finaledit{$z_{\rm s}$} & $\Sigma_{10}$ &    $\Sigma_{10,\text{fs}}$ &    $\Sigma_{20}$ &     $\Sigma_{20,\text{fs}}$ & centroid offset  & $\;\Delta \phi\;$ \\
     & & &(Mpc$^{-2}$ ) & (Mpc$^{-2}$ ) & (Mpc$^{-2}$ ) & (Mpc$^{-2}$ ) & (kpc)  & (\degr)  \\
\hline

\finaledit{J023.0157$-$16.0040} &             $0.36 \pm 0.04$ & 
-- &
\finaledit{$647 \pm 99$} &            
\finaledit{$56 \pm 9$} &             
\finaledit{$853 \pm 132$} &             \finaledit{$82 \pm 13$} &            $0.73 \pm 0.07$ &             $4.1 \pm 2.5$ & \\ 
\finaledit{J024.1631$+$00.1384} &             \finaledit{$0.34\pm0.04$\tablefootmark{a}} & 
\finaledit{$2.628$} &
\finaledit{$389 \pm 56$} &             \finaledit{$61 \pm 9$} &             
\finaledit{$533 \pm 77$} &             \finaledit{$82 \pm 12$} &            $6.09 \pm 0.42$ &             $59.5 \pm 0.4$ & \\ 
\finaledit{J030.4360$-$27.6618} &             $0.75 \pm 0.07$ & 
-- &
\finaledit{$1130 \pm 77$} &             \finaledit{$876 \pm 59$} &             \finaledit{$1245 \pm 85$} &             \finaledit{$723 \pm 49$} &            $0.39 \pm 0.06$ &             $44.8 \pm 12.6$ & \\ 
\finaledit{J033.8095$-$29.1570} &             $0.94 \pm 0.12$ &
-- &
\finaledit{$622 \pm 48$} &             \finaledit{$293 \pm 23$} &             \finaledit{$706 \pm 55$} &             \finaledit{$316 \pm 25$} &            $0.45 \pm 0.04$ &             $9.9 \pm 0.8$ & \\ 
\finaledit{J094.5639$+$50.3059} &             \finaledit{$0.52 \pm 0.08$\tablefootmark{a}} &
-- &
\finaledit{$1431 \pm 257$} &             \finaledit{$119 \pm 21$} &             \finaledit{$718 \pm 126$} &             \finaledit{$137 \pm 24$} &            $1.76 \pm 0.14$ &             $52.0 \pm 0.5$ & \\ 
\finaledit{J140.8110$+$18.4954} &             $0.68 \pm 0.05$ &
\finaledit{$2.417$} &
\finaledit{$455 \pm 30$} &             \finaledit{$226 \pm 15$} &             \finaledit{$400 \pm 26$} &             \finaledit{$242 \pm 16$} &            $0.27 \pm 0.04$ &             $66.1 \pm 19.6$ & \\ 
\finaledit{J154.6972$-$01.3590} &             $0.40 \pm 0.03$ &
\finaledit{$1.432$} &
\finaledit{$1011 \pm 92$} &             \finaledit{$322 \pm 29$} &             \finaledit{$997 \pm 90$} &             \finaledit{$127 \pm 12$} &            $0.97 \pm 0.04$ &             $8.2 \pm 0.4$ & \\ 
\finaledit{J165.4754$-$06.0423} &             \finaledit{$0.33 \pm 0.04$\tablefootmark{a}} & 
\finaledit{$1.675$} &
\finaledit{$1568 \pm 284$} &             \finaledit{$168 \pm 31$} &             \finaledit{$1482 \pm 276$} &             \finaledit{$249 \pm 46$} &            $0.16 \pm 0.02$ &             $1.1 \pm 0.6$ & \\ 
\finaledit{J181.3974$+$41.1790} &             $0.62 \pm 0.02$ &
-- &
\finaledit{$514 \pm 16$} &             \finaledit{$234 \pm 8$} &             \finaledit{$514 \pm 16$} &             \finaledit{$223 \pm 7$} &            $1.50 \pm 0.10$ &             $11.7 \pm 0.8$ & \\ 
\finaledit{J225.4050$+$52.1417} &             $0.75 \pm 0.07$ &  
-- &
\finaledit{$765 \pm 58$} &             \finaledit{$547 \pm 43$} &             \finaledit{$745 \pm 57$} &             \finaledit{$683 \pm 52$} &            $1.74 \pm 0.14$ &             $30.9 \pm 4.0$ & \\ 
\finaledit{J234.4783$+$14.7232} &             \finaledit{$0.65 \pm 0.04$\tablefootmark{a}} & 
-- &
\finaledit{$425 \pm 24$} &             \finaledit{$312 \pm 18$} &             \finaledit{$556 \pm 32$} &             \finaledit{$454 \pm 26$} &            $0.58 \pm 0.18$ &             $13.4 \pm 1.4$ & \\ 
\finaledit{J238.5690$+$04.7276} &             $0.59 \pm 0.06$ &
\finaledit{$1.721$} &
\finaledit{$653 \pm 64$} &             \finaledit{$195 \pm 19$} &             \finaledit{$843 \pm 83$} &             \finaledit{$270 \pm 26$} &            $0.15 \pm 0.03$ &             $5.9 \pm 2.4$ & \\ 
\finaledit{J246.0062$+$01.4836} &             $0.79 \pm 0.05$ &
-- &
\finaledit{$788 \pm 34$} &             \finaledit{$534 \pm 23$} &             \finaledit{$819 \pm 35$} &             \finaledit{$363 \pm 16$} &            $1.13 \pm 0.04$ &             $6.2 \pm 0.9$ & \\ 
\finaledit{J257.4348$+$31.9046} &             $0.72 \pm 0.08$ &
-- &
\finaledit{$252 \pm 25$} &             \finaledit{$148 \pm 14$} &             \finaledit{$349 \pm 34$} &             \finaledit{$168 \pm 16$} &            $0.22 \pm 0.04$ &             $10.4 \pm 0.6$ & \\ 
\finaledit{J329.6820$+$02.9584} &             $0.28 \pm 0.02$ &
\finaledit{$2.080$} &
\finaledit{$1393 \pm 140$} &             \finaledit{$299 \pm 30$} &             \finaledit{$1489 \pm 149$} &             \finaledit{$88 \pm 9$} &            $0.66 \pm 0.04$ &             $10.2 \pm 0.6$ & \\ 
\hline

\end{tabular}
\endgroup

\tablefoot{The spectroscopic redshifts $z_{\rm s}$ of the sources are quoted from \citet{Tran22, bib3:Huang25}. The different definitions of the local galaxy density, denoted as $\Sigma$, are as follows: $\Sigma_{10}$ refers to the projected local galaxy density calculated based on the nearest 10 neighbors; $\Sigma_{10,\text{fs}}$ is similar to $\Sigma_{10}$, but it specifically considers the nearest 10 neighbors selected based on having fluxes greater than 1\% of the corresponding central deflector's flux; $\Sigma_{20}$ and $\Sigma_{20,\text{fs}}$ follow the same definitions as above, but they take into account the 20 nearest neighbors instead of 10. \\
\tablefoottext{a}{\finaledit{Although we used the uniformly available photometric redshifts tabulated above in our analysis, for reference, we provide here the spectroscopic redshifts reported in the literature for J024.1631$+$00.1384: $z_{\rm d} = 0.344$, J094.5639$+$50.3059: $z_{\rm d} = 0.522$,
J165.4754$-$06.0423: $z_{\rm d} = 0.483$, and
J234.4783$+$14.7232: $z_{\rm d} = 0.478$
\citep{Tran22, bib3:Huang25}.}}}
\end{centering}
\end{table*}

\begin{figure*}[htbp]
    \centering
    \begin{tabular}{cc}
        \includegraphics[width=0.45\textwidth]{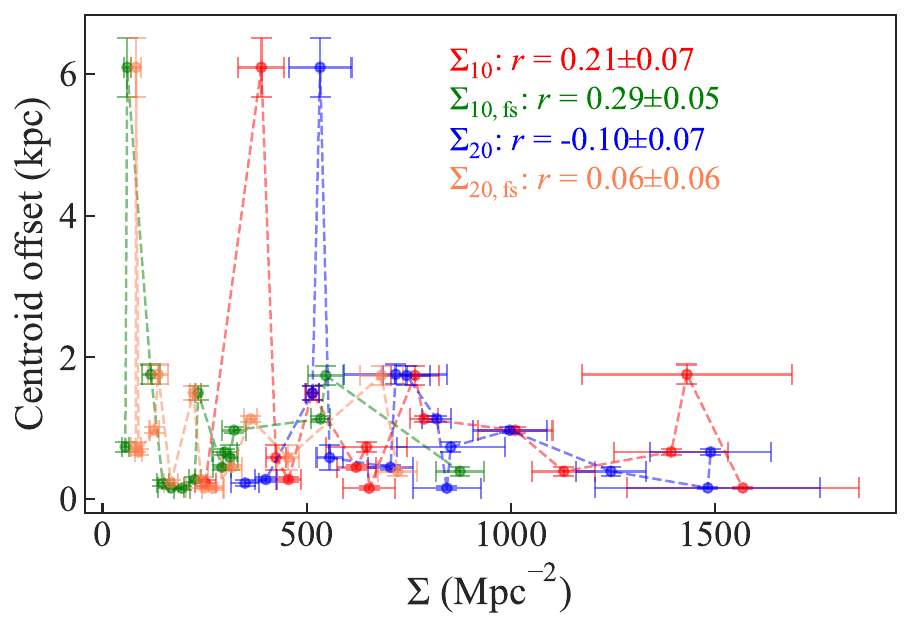} &
        \includegraphics[width=0.45\textwidth]{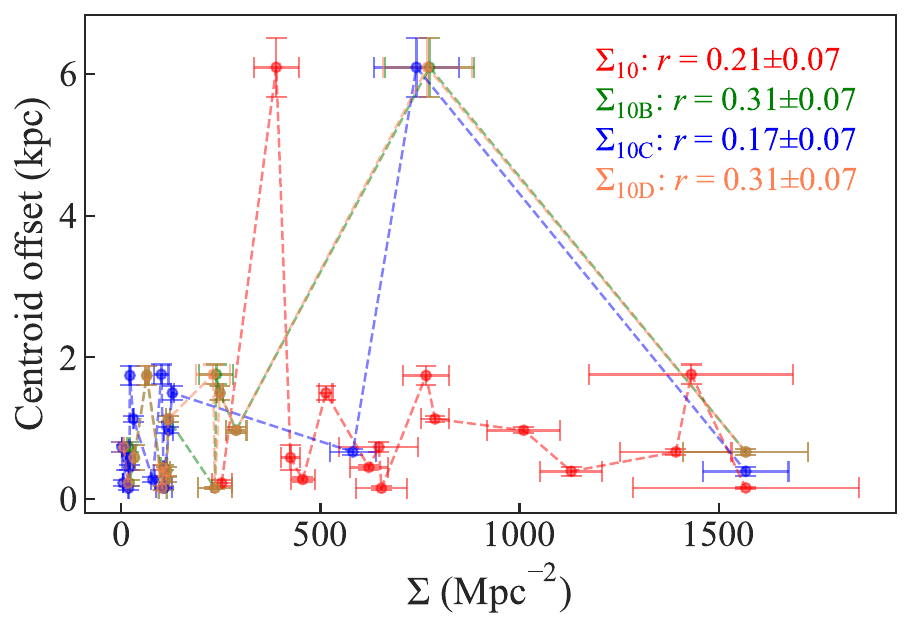}
    \end{tabular}
    \caption{Distribution of the centroid offset between mass and light and the local galaxy density $\Sigma$. The left-hand panel shows the distributions for variants of $\Sigma_{10}$ based on a number of galaxies chosen and applying a flux selection. The right-hand panel illustrates the distributions for various weighting schemes in the $\Sigma$ definition (i.e., definitions B, C, and D). The very weak to weak correlations for all the definitions of $\Sigma$ robustly point out that the local galaxy density has no impact on the centroid offset between the mass and light.}
    \label{fig:cent_off_Sigma}
\end{figure*}

\subsection{Estimation of local galaxy densities} \label{sec:local_densities}

We estimate the local galaxy density using the full field of view in the HST imaging. After background subtraction, we identified all the objects in the image (i.e., galaxies and stars) using the \texttt{detect\_sources} functionality from the image segmentation of \textsc{photutils}\footnote{\url{https://photutils.readthedocs.io/en/stable/segmentation.html}}, which is an affiliated package of \textsc{Astropy}.

To address blending between multiple objects, we employed the \texttt{deblend\_sources} method utilizing multiple thresholds, specifically setting the threshold 3 to 5 times the background RMS level.

We started with threshold 5, and if the threshold did not detect the probable faint sources, we tuned the threshold down to 3. For this lower threshold, the algorithm detects a few artifacts as source objects. After identification, we manually excluded the central deflector galaxy, parts of the lensing arcs, stars, and the remaining artifacts by checking their morphology and central brightness to retain only the neighboring galaxies. We also excluded objects with a high redshift difference with the central deflector. To consider nearby line-of-sight galaxies to be neighbors, we adopted the same criterion used in \citep{Treu09}, which is given by $z_{\text {lens }}-\delta z_{\text {lens }}<z<z_{\text {lens }}+\delta z_{\text {lens }}$, with $\delta z_{\text {lens }}=0.03\left(1+z_{\text {lens }}\right)$. We obtained the photometric redshifts from the DESI Legacy Imaging Surveys DR8 \citep{Duncan22}. We excluded only the galaxies using this photo-$z$ criterion for which the photometric redshifts are available from the DESI Legacy Imaging Surveys.

To quantify how crowded the environment surrounding a lensing galaxy is, we used the local galaxy density \citep{Dressler80, Cooper05}. A common parameter (with high dynamic range) used in this area of investigation is ``$n^{\rm th}$ nearest-neighbor density'' \citep{Treu09}. It is defined as the density (of galaxies) considering the nearest $n$ neighbors of the central deflector. We assess common environmental factors as outlined in works such as \citet{Cooper05}. Specifically, we determined the galaxy density within a circle whose radius matches the distance to the lens's tenth closest neighboring galaxy, denoted as $\Sigma_{10}$ \citep[as described by ][]{Dressler80}. This  $\Sigma_{10}$ is the baseline galaxy density used in this analysis. To test the robustness of our findings against the specific definition of $\Sigma_{10}$, we alternatively computed this density using the twentieth nearest neighbor, labeled $\Sigma_{20}$. Additionally, we also adopted a variation of these  $\Sigma_{10}$ and  $\Sigma_{20}$ quantities by excluding some galaxies based on the flux values. In these flux-selected variations (labeled $\Sigma_{10, \rm fs}$ or $\Sigma_{20, \rm fs}$), galaxies with flux less than that of the $1\%$ of the central deflecting galaxy were excluded. The motivation behind this flux selection is that the galaxies with very low fluxes (and thus low masses) have a relatively small gravitational effect on other galaxies in the environment. \newedit{Additionally, this variation excludes sources that do not have photo-$z$ estimates from the DESI Legacy Imaging Surveys. Therefore, these flux-selected variations of $\Sigma$ would lead to better purity when selecting galaxies that belong to the environment of the central deflector.} \newedit{Conversely, for the flux-selected variations, the radius covering the \nth{10} or \nth{20} neighbor becomes much larger than their `vanilla' counterpart, thus diluting the interpretability of the $\Sigma$ values as the `local' galaxy density.} Furthermore, the four definitions above are limited in their scope as they do not incorporate any information about the sizes or distances of the neighbors from the center. To check any potential systematics arising from the rather simple definition of the local galaxy density, we devise three extended definitions as follows (definition A corresponds to the definitions given above. However, we choose to omit the subscript `A' in the notation for conciseness):
\begin{itemize}
    \item Definition B: This definition considers the total mass (inferred from the total flux) inside the $n$-th neighbor radius. Individual fluxes are added up, and the sum is normalized by the global mean flux, which is calculated from all 15 systems.
    \\
    \item Definition C: This definition incorporates both the mass and distance from the central deflector for each of the neighboring galaxies. Each mass is weighted by the inverse of the distance, and the total sum is normalized by the global mean of the flux-over-distance quantities.
    \\
    \item Definition D: As distance is in the denominator of definition C, the terms blow up when considering galaxies very close to the central deflector. To avoid such divergences, a constant weighting of 66 kpc is used for all the galaxies inside 66 kpc. This 66 kpc is the physical distance corresponding to 10 arcsec angular distance at the mean deflector redshift ($\langle z \rangle \sim$0.581) for our sample of 15 systems. The galaxies beyond this distance are weighted as in definition C. Similar weighting schemes are discussed in \citet{Greene13} for defining weighted number counts as a metric of galaxy overdensity.
\end{itemize}
 
\finaledit{We calculated the uncertainties for the local galaxy density quantities ($\Sigma$) by propagating the redshift uncertainties of the line-of-sight galaxies using Monte Carlo sampling. For a sanity check, we also computed the uncertainties using analytic formulae for uncertainty propagation, which yielded similar uncertainty values.} We tabulate the values for $\Sigma_{10}$, $\Sigma_{10, \rm fs}$, $\Sigma_{20}$, and $\Sigma_{20, \rm fs}$ in Table \ref{table:photometric_params}. We present the correlations of these quantities with the offset or misalignment between the mass and light next in Section \ref{sec:alignment_results}.

\begin{figure*}[htbp]
    \centering
    \begin{tabular}{cc}
        \includegraphics[width=0.45\textwidth]{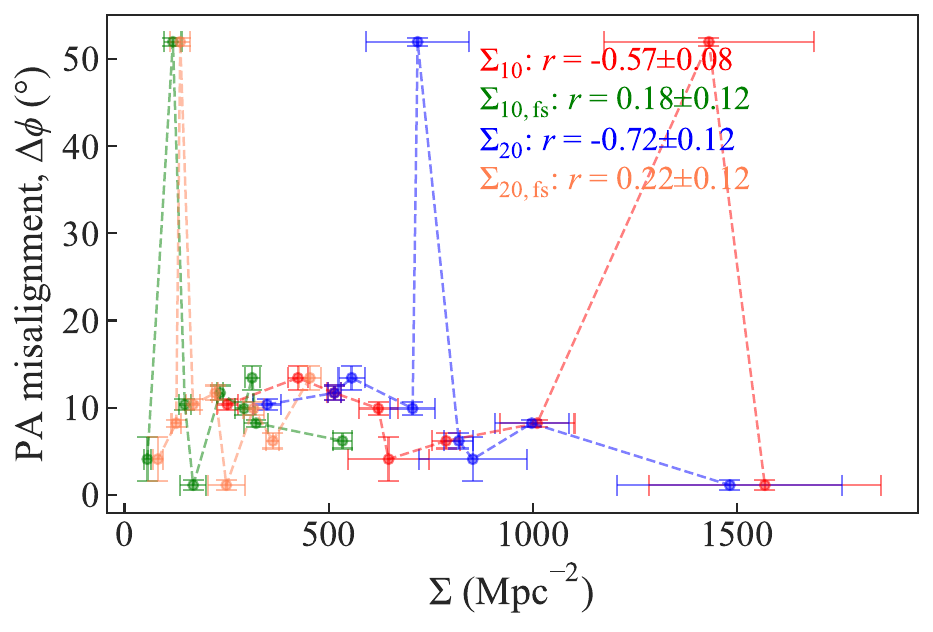} &
        \includegraphics[width=0.45\textwidth]{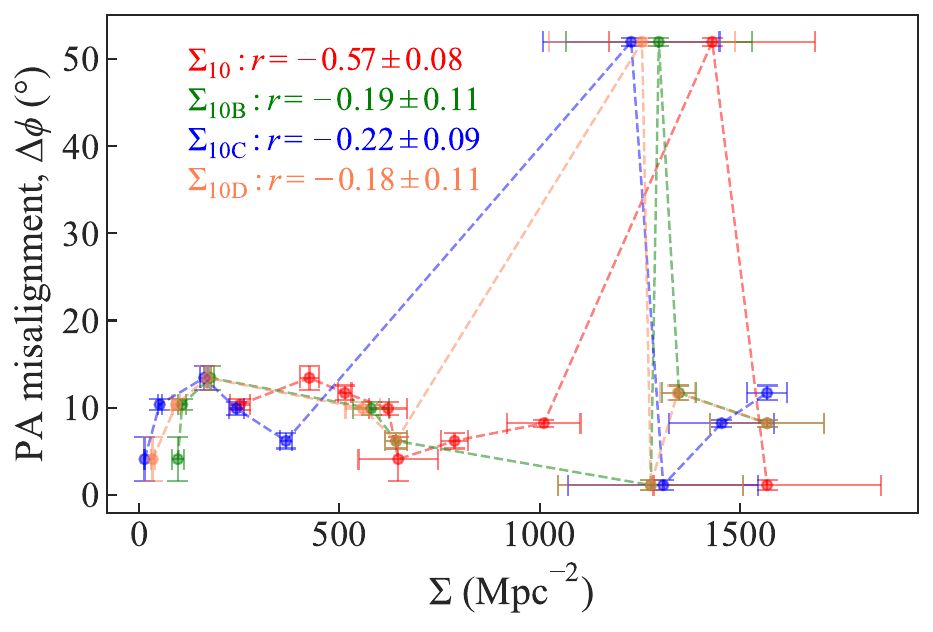}
    \end{tabular}
    \caption{\newedit{
    The distribution of the PA misalignment between the mass and light, and the local galaxy density $\Sigma_{10}$ and its variations. The variations of $\Sigma_{10}$ in the left-hand panel are based on different galaxy selection criteria, including the number of galaxies considered and flux-based thresholds. The variations in the right-hand panel correspond to different weighting schemes used in defining $\Sigma_{10}$ (i.e., definitions B, C, and D). Systems with low ellipticity ($q_\text{L} > 0.9$) are excluded from both cases. \finaledit{While the baseline definitions $\Sigma_{10}$ and $\Sigma_{20}$ exhibit a moderate to strong correlation with the PA misalignment, other definitions show only weak to very weak correlations.} Therefore, we cannot robustly conclude a strong relationship between the PA misalignment and the local galaxy density.}
    }
    \label{fig:pos_off_Sigma}
\end{figure*}

\subsection{Mass and light alignment} \label{sec:alignment_results}

In this subsection, we present our results on the alignment between the mass and light distributions in our sample of lensing systems.

\subsubsection{Centroid}
In this study, we adopt the bi-weight mid-correlation (hereafter, correlation) as the correlation measure between the quantities of our interest.
The correlation between the centroid offset and the local galaxy density is weak ($r=0.21 \pm \finaledit{0.07}$ for the baseline definition $\Sigma_{10}$). The correlation remains between weak to very weak for three other alternative definitions of local galaxy density incorporating different numbers of nearest neighbors and minimum flux level (Fig. \ref{fig:cent_off_Sigma}, left panel). Similarly, the alternative definitions B, C, and D with different flux and distance-based weighting schemes for the $\Sigma$ quantity also give a similar result (very weak to weak correlation) (Fig. \ref{fig:cent_off_Sigma}, right panel). Therefore, we robustly find no impact of the lens galaxy's environment on the offset between the mass and light centroids.

\subsubsection{Position angle} \label{sec:pa_correlations}

In investigations involving the PA of the major axis, the axis ratio of light and mass profiles of the considered system should be small enough for the major axes to be well-defined \citep{Treu09}. Therefore, when computing the associated correlations involving the PA misalignment, we excluded six out of the 15 systems that have a high light-axis ratio (i.e., $q_\text{L}>0.9$). The correlation between the PA misalignment and the local galaxy density is \finaledit{moderate ($r= -0.57 \pm 0.08$ for the baseline definition $\Sigma_{10}$). The correlation becomes strong for the alternative definition $\Sigma_{20}$}, but for the definition with setting minimum flux level, the correlation becomes very weak to weak (Fig. \ref{fig:pos_off_Sigma}, left panel). For the $\Sigma$ definitions B, C, and D, the correlation stays \finaledit{very weak to weak} (Fig. \ref{fig:pos_off_Sigma}, right panel). As a result, we cannot robustly conclude that there is a strong correlation between the PA misalignment and the local galaxy density, as the level of correlation is dependent upon the definition of the local galaxy density.

\begin{figure*}[htbp]
    \centering
    \begin{tabular}{cc}
        \includegraphics[width=0.45\textwidth]{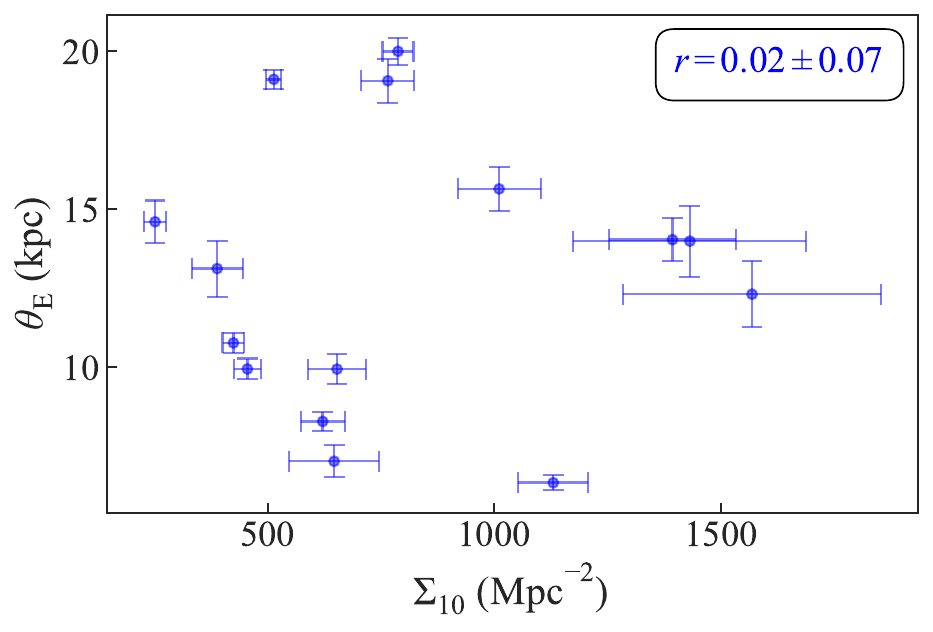} & 
        \includegraphics[width=0.45\textwidth]{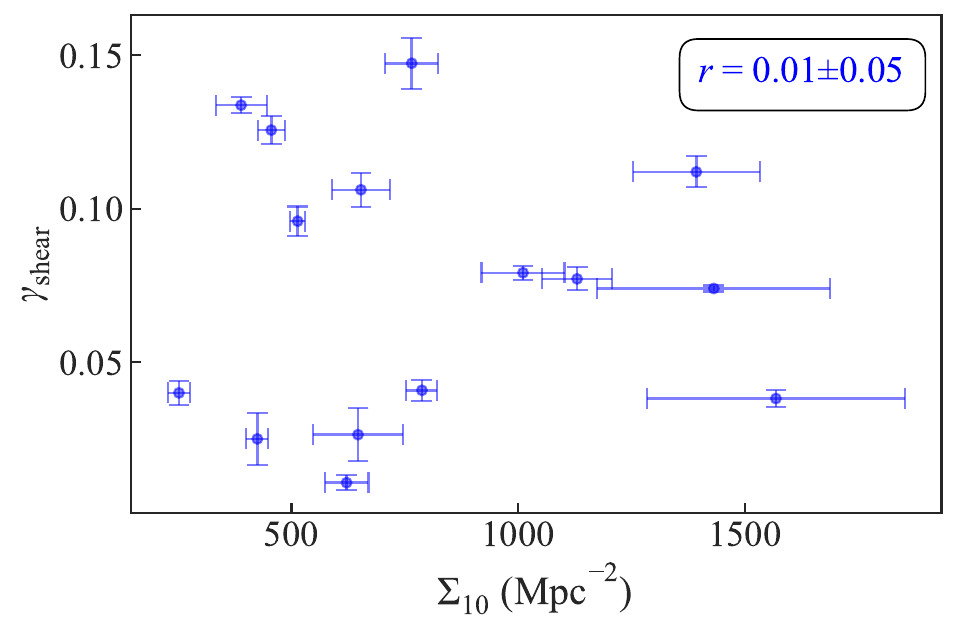}
    \end{tabular}
    
    \caption{\newedit{The distribution of the local galaxy density $\Sigma_{10}$ vs. the Einstein radius \finaledit{$\theta_{\rm E}$} (left-hand panel) and vs. the residual shear magnitude \finaledit{$\gamma_{\rm shear}$} (right-hand panel). The correlation is very weak for both the Einstein radius and the residual shear magnitude.}}
    \label{fig:einstein_radius_gamma_shear_Sigma_10A}
\end{figure*}

\subsection{Correlation between the local galaxy density and lens model parameters}
\label{sec:lens_model_params}

\begin{figure}[htbp]
  \centering
  \includegraphics[width=0.45\textwidth]{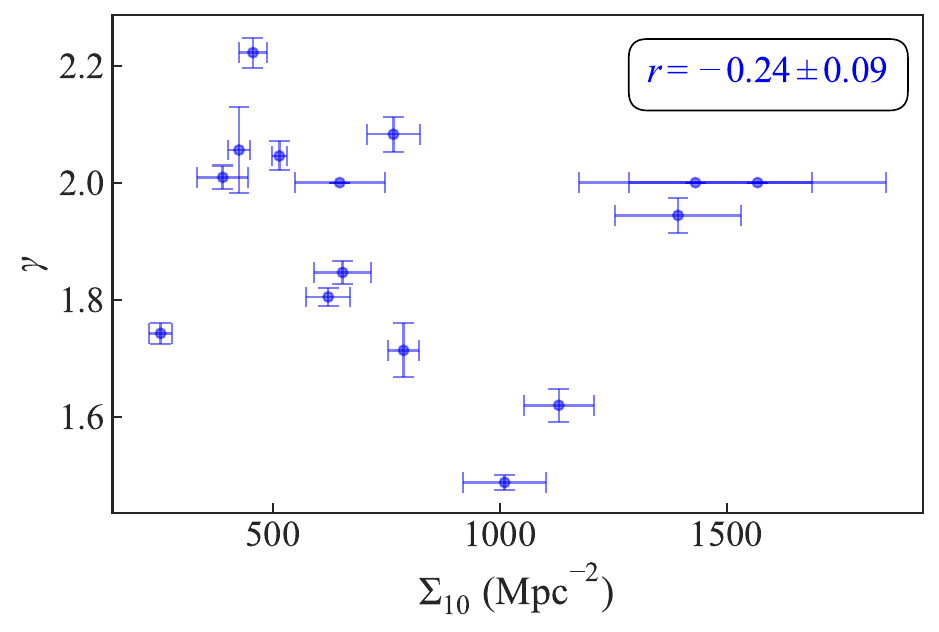}
  \caption{\newedit{ The logarithmic slope $\gamma$ vs. the distribution of the local galaxy density $\Sigma_{10}$. We find only a weak correlation between the two, which is consistent with \citet{Treu09}.}}
  \label{fig:gamma_Sigma_10A}
\end{figure}

\newedit{As illustrated in Fig. \ref{fig:einstein_radius_gamma_shear_Sigma_10A} (left-hand panel), we find the correlation between the Einstein radius and the local galaxy density is very weak ($r= 0.02 \pm \finaledit{0.07}$ for the baseline definition $\Sigma_{10}$). The correlation stays very weak between the residual shear magnitude, $\gamma_{\rm shear}$ and the local galaxy density giving $r= 0.01 \pm \finaledit{0.05}$ for the baseline definition $\Sigma_{10}$ (Fig. \ref{fig:einstein_radius_gamma_shear_Sigma_10A}, right panel). The logarithmic slope $\gamma$ is also weakly ($r= \finaledit{-0.24 \pm 0.09}$) correlated with the local galaxy density (Fig. \ref{fig:gamma_Sigma_10A}), consistent with the finding of \citet{Treu09}.} \newedit{If a correlation were detected, it would support the hypothesis that the local slope of the mass tends to be steeper in a more crowded environment, possibly as the result of tidal truncation as indicated by some simulations \citep[e.g.,][]{Dobke07}. Therefore, our result either indicates that this hypothesis is incorrect, or at least the predicted effect is not present at kpc scales, or the environmental contribution to the projected mass density that we have not explicitly distinguished in our model is sufficient to negate the signature of the effect.}

\begin{figure}[htbp]
  \centering
  \includegraphics[width=0.45\textwidth]{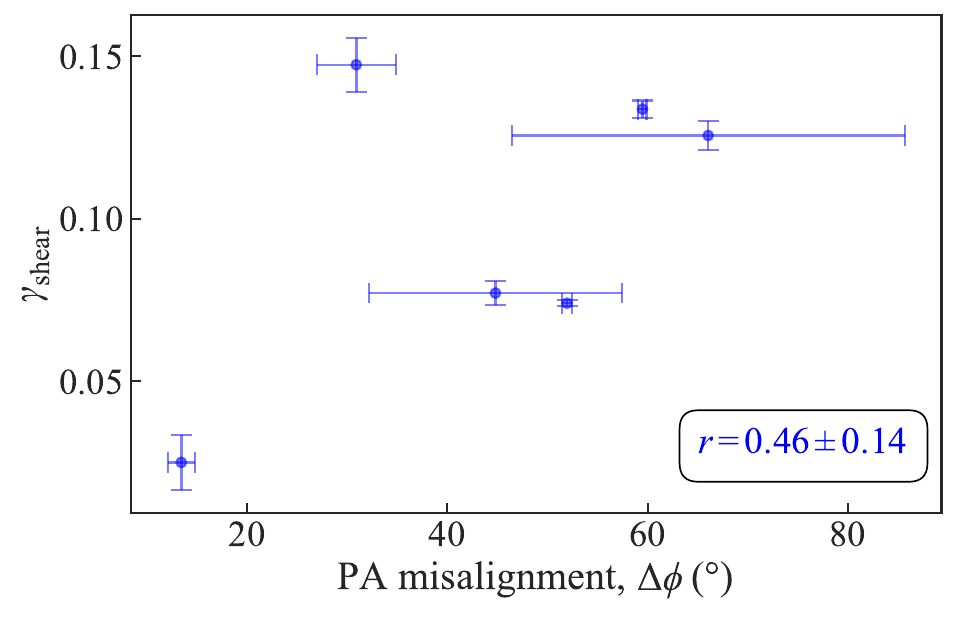}
  \caption{\newedit{The residual shear magnitude \(\gamma_{\rm shear}\) vs. the distribution of the PA misalignment $\Delta \phi$. These two quantities are moderately correlated ($r=\finaledit{0.46\pm14})$. Here, we excluded systems with low PA misalignment, that is, those with $\Delta \phi \lesssim 10 \degr $.}}
  \label{fig:gamma_shear_vs_pa_offset}
\end{figure}

\newedit{We see a moderate correlation ($r= \finaledit{0.46 \pm 0.14}$) between the residual shear magnitude and the PA misalignment (excluding systems with low misalignment, i.e., $\Delta \phi \lesssim 10 \degr $), which is illustrated in Fig. \ref{fig:gamma_shear_vs_pa_offset}. 
This result is in agreement with several previous studies \citep[e.g.,][]{Gavazzi12, Bruderer16, Shajib19}}. Here, we did not apply the $q_{\rm L} > 0.9$ cut as applied in Section \ref{sec:pa_correlations}, since this additional cut on top of the  \finaledit{$\Delta\phi \gtrsim 10\degr$} cut would not leave enough systems to obtain a robust correlation.

\section{Discussion} 
\label{sec:discussion}

A majority (9 out of 15 systems) of the PA misalignment values are $\lesssim 10^{\degr}$, which is consistent with several previous studies \citep{Keeton98b, Kochanek02, Treu09, Gavazzi12, Sluse12, Bruderer16, Shajib19, Shajib21}. \newedit{Our observation of the moderate correlation between the residual shear magnitude and the PA misalignment for cases with $\Delta \phi \gtrsim 10\degr$ also agrees with previous findings in the literature \citep[e.g.,][]{Bruderer16, Shajib19}. }

For the regular definitions of the local galaxy density (i.e., $\Sigma_{\rm 10}$ and $\Sigma_{\rm 20}$), we find \finaledit{a moderate to strong} correlation between them and the PA misalignment (Fig. \ref{fig:pos_off_Sigma}). This result agrees very well with \citet{Treu09}, who also found a strong correlation between the PA misalignment and $\Sigma_{10}$. However, for alternative definitions of $\Sigma$ that apply a flux selection on the galaxies or weigh the galaxies differently based on their flux and distance from the central deflector, this correlation becomes weak to very weak (Fig. \ref{fig:pos_off_Sigma}). This either calls into question the dependency of this strong correlation on the particular definition of local galaxy density \newedit{or the applicability of the alternative definitions of $\Sigma$ to be considered a reliable metric of the local galaxy density despite their intended design.} 

However, the weakened support for the environmental origin of the PA misalignment, and our finding of no correlation between the residual shear magnitude and the local galaxy density, paints a coherent picture with the argument presented by \citet{Etherington23}. These authors argued that the residual shear magnitude is not solely due to external structures as suggested by the commonly used terminology `external shear.' Rather, the necessity of including a residual shear field in the lens model, a widely common practice in the literature, stems from the inadequacy of a single, uniformly elliptical mass distribution to capture the angular complexities truly present in the central deflector galaxies. Since we find the larger values of the PA misalignment angle to be moderately correlated with the residual shear magnitude, such large PA misalignments can also originate from the model having to compensate for potentially unaccounted angular complexities in the deflector. In that case, local galaxy density is not expected to correlate with the PA misalignment since the latter observable does not have a physical origin. More sophisticated models that account for deflector angular complexity to a greater detail \citep[e.g.,][]{He24, Amvrosiadis24} will be necessary to robustly settle the connection between the `true' PA misalignment and the local environment.

We find the mean centroid offset in our sample of 15 lenses to be $0.42 \pm 0.01$ kpc. 
\citet{Schaller15} found the centroid offset between the dark and luminous matter in galaxies from the EAGLE simulation to be well described by a Maxwellian distribution with $\sigma = 0.2$ kpc. Therefore, our mean is situated at the $\sim$\nth{75} percentile of that distribution and thus well within the 2$\sigma$ consistency or the \nth{95} percentile. However, a caveat in this comparison is that the simulation results were based on the offset between dark matter and luminous matter centroids. In contrast, our centroid offsets are between the total matter and luminous matter. To facilitate a more robust comparison between simulations and observations, either the same observable needs to be extracted from the simulated galaxies, or lens models where dark matter and luminous matter distributions are separately accounted for \citep[e.g.,][]{Suyu14, Shajib20, Shajib22b} would be necessary. Both of these explorations are out of the scope of this paper, and we leave them for future investigations.

Among previous studies, \citet{Shajib21} constrained a 68 percent upper limit of 218$\pm$19 pc from a sample of 23 galaxy--galaxy lenses, which is consistent with our result within 2$\sigma$. As we robustly find the centroid offset to be non-correlated with the local galaxy density, we provide support for using the centroid offset as a robust probe of dark matter theories \citep{Harvey14, Kahlhoefer14, Robertson17}.

The considered sample contains an outlier system (\finaledit{DESI J024.1631$+$00.1384}) with a centroid offset of $6.1\pm0.4$ kpc. On visual inspection,
it appears to be a merger of two galaxies. In similar previous studies, the only observed system with a significantly larger offset ($1.72\pm 0.42$ kpc) is a merging galaxy system \citep{Shu16}. Although there was an initial report of a similar offset for a cluster-member galaxy in Abell 3827 \citep{Williams11, Massey15}, the offset was ruled out with more data \citep{Massey18}. \citet{Shajib19} also identified an outlier in their sample of 13 lensing systems that similarly has two comparable-mass deflectors in close proximity.

\finaledit{A useful extension for similar studies in the future but with larger samples would be to categorize the sample into central and satellite galaxies and compare the internal structures between the two classes. As the central and satellites can have different evolutionary histories or formation timelines \citep{Simha09}, and are impacted in different ways through their interactions with the environment \citep{Zhu24}, any differences in their structural properties, or the lack thereof \citep{Wang18}, could offer valuable insights into the nature of those interactions in these two populations. However, due to our sample size, dividing it into smaller subsamples is unlikely to yield statistically meaningful results in this regard. Therefore, we leave such an investigation for future studies incorporating larger samples.}

\section{Summary}
\label{sec:summary}

In this paper, we modeled a collection of 15 galaxy--galaxy gravitational lensing systems. These lens models are the first to be presented in the literature for these systems. In the modeling process, we used the publicly available lens modeling software package \textsc{lenstronomy}. Our baseline model for the lens mass comprises an EPL profile for the central deflector galaxy and a residual shear field. We fine-tuned this baseline mass model as necessary for more complex systems in the sample, for example, explicitly including nearby satellites and companions and accounting for higher-order lensing perturbation with a flexion field. For several systems, it was necessary to adjust the masks used when fitting the imaging data to reach a highly optimal stage. Additionally, for some systems, it was also required to adopt extra source components when needed. 

We measured the local galaxy density $\Sigma_{10}$ for all our systems and investigated its correlation with the centroid offset and PA misalignment between the mass and light. We also tested our results on these correlations against alternative definitions of the local galaxy density. Our main results are:

\begin{itemize}
    \item We find no impact of the local environment on the offset between the light and mass centroids, i.e., the correlation is robustly identified as weak to very weak between these two.

    \item We find \finaledit{a moderate to strong} correlation between the PA misalignment and the standard definitions of the local galaxy density (i.e., $\Sigma_{\rm 10}$ and $\Sigma_{\rm 20}$), which agrees very well with previous study \citep{Treu09}. However, this finding of \finaledit{moderate to strong} correlation is not robust against changing the definitions of the local galaxy density.
\end{itemize}

This paper ventured into modeling group-scale systems, extending the methodology ubiquitously applied to galaxy-scale ones. Group-scale systems are naturally more complex to model due to the need to account for multiple nearby companions and satellites than the case of a galaxy-scale system with a single main deflector \citep[with one or two small satellites for $\lesssim$30\% cases, e.g.,][]{Shajib19}. These group-scale lenses provide a useful probe of the connection between galaxies' internal structure and the environment, as investigated in this paper. However, modeling such complex systems can require a high level of fine-tuning in the model setup, as we have encountered in this analysis, which is time-consuming for the investigators. Various large-area sky surveys such as Euclid, the Vera Rubin Observatory Legacy Survey of Space and Time, and the Roman Space Telescope are expected to find strong lensing systems, including group-scale ones, in unprecedentedly large numbers, increasing the catalog size of currently known lenses by two orders of magnitudes or more \citep{Oguri10, Collett15, Shajib24b}. Automated modeling pipelines, whether through tailored algorithms \citep[e.g.,][]{Shajib19, Shajib21, bib3:Schmidt23} or through a fully machine-learning-based approach \citep[e.g.,][]{Erickson24} will be highly advantageous to avoid a huge requirement of investigator time when modeling a large sample of group-scale lenses.

\begin{acknowledgements}
    This paper is the product of an \href{https://www.astrobridge.org/}{Astro Bridge} research project titled \href{https://www.astrobridge.org/projects/bdlensing}{BD Lensing} (BD is the two-letter abbreviation of Bangladesh). This project was offered as a bridge program to undergraduate, post-baccalaureate, and Master's students in Bangladesh aspiring to pursue a PhD in astrophysics or related fields. AJS mentored and coordinated the project. The student participants who provided substantial contributions to this paper were deemed co-authors. The first five authors (SMRA, MJH, AAI, SHR, and FRS) are considered joint first authors since they are equally top contributors to all aspects of this project and manuscript writing. The joint first authors can prioritize their own name when including this paper in their curriculum vitae. \\

      AJS thanks Khan Muhammad Bin Asad for helping to recruit students for the BDLensing project. Support for this work was provided by NASA through the NASA Hubble Fellowship grant HST-HF2-51492 awarded to AJS by the Space Telescope Science Institute, which is operated by the Association of Universities for Research in Astronomy, Inc., for NASA, under contract NAS5-26555.\\

      This article made use of \textsc{lenstronomy} \citep{bib3:Birrer18, bib3:Birrer21},  \textsc{emcee} \citep{Foreman-Mackey13}, \textsc{astropy} \citep{bib3:AstropyCollaboration13, bib3:AstropyCollaboration18, bib3:Astropy22}, \textsc{numpy} \citep{bib3:Oliphant15}, \textsc{scipy} \citep{bib3:Jones01}, \textsc{matplotlib} \citep{bib3:Hunter07}, and \textsc{jupyter} \citep{Kluyver16}.
\end{acknowledgements}

\bibliographystyle{aa}
\bibliography{ajshajib}

\begin{appendix}

\begin{table}[h!]
\onecolumn
\section{Model summary} \label{app:models}
\finaledits{Here in Table \ref{table:lens_profiles}, we summarize the model setups for each of the 15 strong lensing systems, as a complement to Section \ref{ssec:specific_descriptions}.}

\begingroup
	\setlength{\tabcolsep}{4pt} 
	\renewcommand{\arraystretch}{1.3} 

\caption{\noindent
Summary for the model setups for the \nlens\ strong lensing systems in our sample.
\label{table:lens_profiles}
}

\begin{tabular}{llllcl}
\hline

System &
  Mass Profiles &
  Lens-light profiles &
  Source-light profiles &
  Additional Priors &
  Modeler(s) \\ \hline
  \hline 
  \begin{tabular}[l]{@{}l@{}}
  \finaledit{DESI J023.0157$-$16.0040} \\ \finaledit{(J0132$-$1600)} \end{tabular} &
  EPL, Shear &
  Double Elliptical S\'ersic &
  \begin{tabular}[l]{@{}l@{}}Elliptical S\'ersic, \\ Shapelets ($n_\text{max} = 10$)\end{tabular} &
  -- &
  \begin{tabular}[c]{@{}l@{}}FRS, \\ AB\end{tabular} \\ \hline
\begin{tabular}[l]{@{}l@{}}\finaledit{DESI J024.1631$+$00.1384} \\ \finaledit{(J0136$+$0008)}\end{tabular} &
  EPL, SIE, Shear &
  \begin{tabular}[l]{@{}l@{}}Double Elliptical S\'ersic, \\ Satellite: Elliptical S\'ersic\end{tabular} &
  \begin{tabular}[l]{@{}l@{}}Elliptical S\'ersic, \\ Shapelets ($n_\text{max} = 10$)\end{tabular} &
  \begin{tabular}[c]{@{}l@{}}$ q_\text{m} > \,q_\text{L} $, \\ $\dfrac{\theta_{\rm E, Cen}}{\theta_{\rm E, Sat}} = \sqrt{\dfrac{\rm flux_{Cen}}{\rm flux_{Sat}}}$  \end{tabular} &
  \begin{tabular}[c]{@{}l@{}}SHR, \\ FRS, \\ AAC\end{tabular} \\ \hline
\begin{tabular}[l]{@{}l@{}}
\finaledit{DESI J030.4360$-$27.6618} \\ \finaledit{(J0201$-$2739)}\end{tabular} &
  EPL, SIE, Shear &
  \begin{tabular}[l]{@{}l@{}}Double Elliptical S\'ersic, \\ Satellite: Elliptical S\'ersic\end{tabular} &
  \begin{tabular}[l]{@{}l@{}}Elliptical S\'ersic, \\ Shapelets ($n_\text{max} = 8$)\end{tabular} &
  \begin{tabular}[c]{@{}l@{}}$ q_\text{m} > \,q_\text{L} $, \\ $\dfrac{\theta_{\rm E, Cen}}{\theta_{\rm E, Sat}} = \sqrt{\dfrac{\rm flux_{Cen}}{\rm flux_{Sat}}}$ \end{tabular} &
  SMRA \\ \hline
\begin{tabular}[l]{@{}l@{}}
\finaledit{DESI J033.8095$-$29.1570} \\ \finaledit{(J0215$-$2909)}\end{tabular}&
  EPL, Shear &
  Double Elliptical S\'ersic &
  \begin{tabular}[l]{@{}l@{}}Elliptical S\'ersic, \\ Double Shapelets \\ ($n_\text{max} = 10$)\end{tabular} &
  $ q_\text{m} > \,q_\text{L} $ &
  NJ \\ \hline
\begin{tabular}[l]{@{}l@{}}
\finaledit{DESI J094.5639$+$50.3059} \\ \finaledit{(J0618$+$5018)}\end{tabular} &
  EPL, SIE, Shear &
  \begin{tabular}[l]{@{}l@{}}Double Elliptical S\'ersic, \\ Satellite: Elliptical S\'ersic\end{tabular} &
  \begin{tabular}[c]{@{}l@{}}Elliptical S\'ersic, \\ Shapelets ($n_\text{max} = 10$)\end{tabular} &
  \begin{tabular}[c]{@{}l@{}}$ q_\text{m} > \,q_\text{L} $, \\ $\dfrac{\theta_{\rm E, Cen}}{\theta_{\rm E, Sat}} = \sqrt{\dfrac{\rm flux_{Cen}}{\rm flux_{Sat}}}$ \end{tabular} &
  \begin{tabular}[c]{@{}l@{}}AAI, \\ AR\end{tabular} \\ \hline
\begin{tabular}[l]{@{}l@{}} 
\finaledit{DESI J140.8110$+$18.4954} \\ \finaledit{(J0923$+$1829)}\end{tabular} &
  EPL, SIE, Shear &
  \begin{tabular}[l]{@{}l@{}}Single Elliptical S\'ersic, \\ Satellite: Elliptical S\'ersic\end{tabular} &
  \begin{tabular}[l]{@{}l@{}}Elliptical S\'ersic, \\ Shapelets ($n_\text{max} = 10$)\end{tabular} &
  -- &
  MR \\ \hline
\begin{tabular}[l]{@{}l@{}} 
\finaledit{DESI J154.6972$-$01.3590} \\ \finaledit{(J1018$-$0121)}\end{tabular} &
  EPL, Shear &
  Single Elliptical S\'ersic &
  \begin{tabular}[l]{@{}l@{}}Elliptical S\'ersic, \\ Shapelets ($n_\text{max} = 10$)\end{tabular} &
  -- &
  MJH \\ \hline
\begin{tabular}[l]{@{}l@{}} 
\finaledit{DESI J165.4754$-$06.0423} \\ \finaledit{(J1101$-$0602)}\end{tabular} &
  EPL, Shear &
  Double Elliptical S\'ersic &
  \begin{tabular}[l]{@{}l@{}}Elliptical S\'ersic, \\ Shapelets ($n_\text{max} = 8$)\end{tabular} &
  -- &
  \begin{tabular}[c]{@{}l@{}}SMRA, \\ JF\end{tabular} \\ \hline
\begin{tabular}[l]{@{}l@{}} 
\finaledit{DESI J181.3974$+$41.1790} \\ \finaledit{(J1205$+$4110)}\end{tabular} & 
  EPL, Shear, Flexion &
  Double Elliptical S\'ersic &
  \begin{tabular}[l]{@{}l@{}}Elliptical S\'ersic, \\ Shapelets ($n_\text{max} = 10$)\end{tabular} &
  -- &
  \begin{tabular}[c]{@{}l@{}}SHR, \\ MSH\end{tabular}  \\ \hline
\begin{tabular}[l]{@{}l@{}} 
\finaledit{DESI J225.4050$+$52.1417} \\ \finaledit{(J1501$+$5208)}\end{tabular} &
  EPL, Shear &
  Double Elliptical S\'ersic &
  \begin{tabular}[l]{@{}l@{}}Elliptical S\'ersic, \\ Shapelets ($n_\text{max} = 8$)\end{tabular} &
  -- &
  MHN \\ \hline
\begin{tabular}[l]{@{}l@{}} 
\finaledit{DESI J234.4783$+$14.7232} \\ \finaledit{(J1537$+$1443)}\end{tabular} &
  EPL, Shear &
  Single Elliptical S\'ersic &
  \begin{tabular}[l]{@{}l@{}}Elliptical S\'ersic, \\ Shapelets ($n_\text{max} = 6$)\end{tabular} &
  -- &
  MHT \\ \hline
\begin{tabular}[l]{@{}l@{}} 
\finaledit{DESI J238.5690$+$04.7276} \\ \finaledit{(J1554$+$0443)}\end{tabular} &
  EPL, Shear &
  Double Elliptical S\'ersic &
  \begin{tabular}[l]{@{}l@{}}Elliptical S\'ersic, \\ Shapelets ($n_\text{max} = 5$)\end{tabular} &
  -- &
  ZJ \\ \hline
\begin{tabular}[l]{@{}l@{}} 
DESI J246.0062$+$01.4836 \\ (J1624$+$0129)\end{tabular} &
  \begin{tabular}[l]{@{}l@{}}EPL,  SIE, Shear\end{tabular} &
  \begin{tabular}[l]{@{}l@{}}Double Elliptical S\'ersic, \\ Satellite: Elliptical S\'ersic\end{tabular} &
  \begin{tabular}[l]{@{}l@{}}Elliptical S\'ersic, \\ Shapelets ($n_\text{max} = 10$)\end{tabular} &
  -- &
  AAI \\ \hline
\begin{tabular}[l]{@{}l@{}} 
DESI J257.4348$+$31.9046 \\ (J1709$+$3154)\end{tabular} &
  EPL, Shear &
  Double Elliptical S\'ersic &
  \begin{tabular}[l]{@{}l@{}}Elliptical S\'ersic, \\ Shapelets ($n_\text{max} = 8$)\end{tabular} &
  -- &
  TA \\ \hline
\begin{tabular}[l]{@{}l@{}} 
DESI J329.6820$+$02.9584 \\ (J2158$+$0257)\end{tabular} &
  EPL, Shear &
  Double Elliptical S\'ersic &
  \begin{tabular}[l]{@{}l@{}}Elliptical S\'ersic, \\ Shapelets ($n_\text{max} = 10$)\end{tabular} &
  -- &
  FRS \\ \hline
\end{tabular}
\endgroup
\tablefoot{In addition to enlisting the system names in DESI's decimal convention, we also provide their alternative sexagesimal version (in parentheses). The description of each model setting is described in Section \ref{sec:lens_modeling}. The contributing modelers for each system are also tabulated. For some systems, we have imposed a prior condition on the satellite galaxy's Einstein radius for it to scale with the mass, assuming the mass-to-light ratio is the same for both the central deflector and the satellite. This condition is denoted with $\theta_{\rm E, Cen}/\theta_{\rm E, Sat} = \sqrt{{\rm flux_{Cen}}/{\rm flux_{Sat}}}$ for the cases it was used.}
\end{table}

\onecolumn
\section{Illustration of lens models} \label{app:model_figures}
\begin{figure*}[h!]
	\centering
	\includegraphics[width=1.2\textwidth]{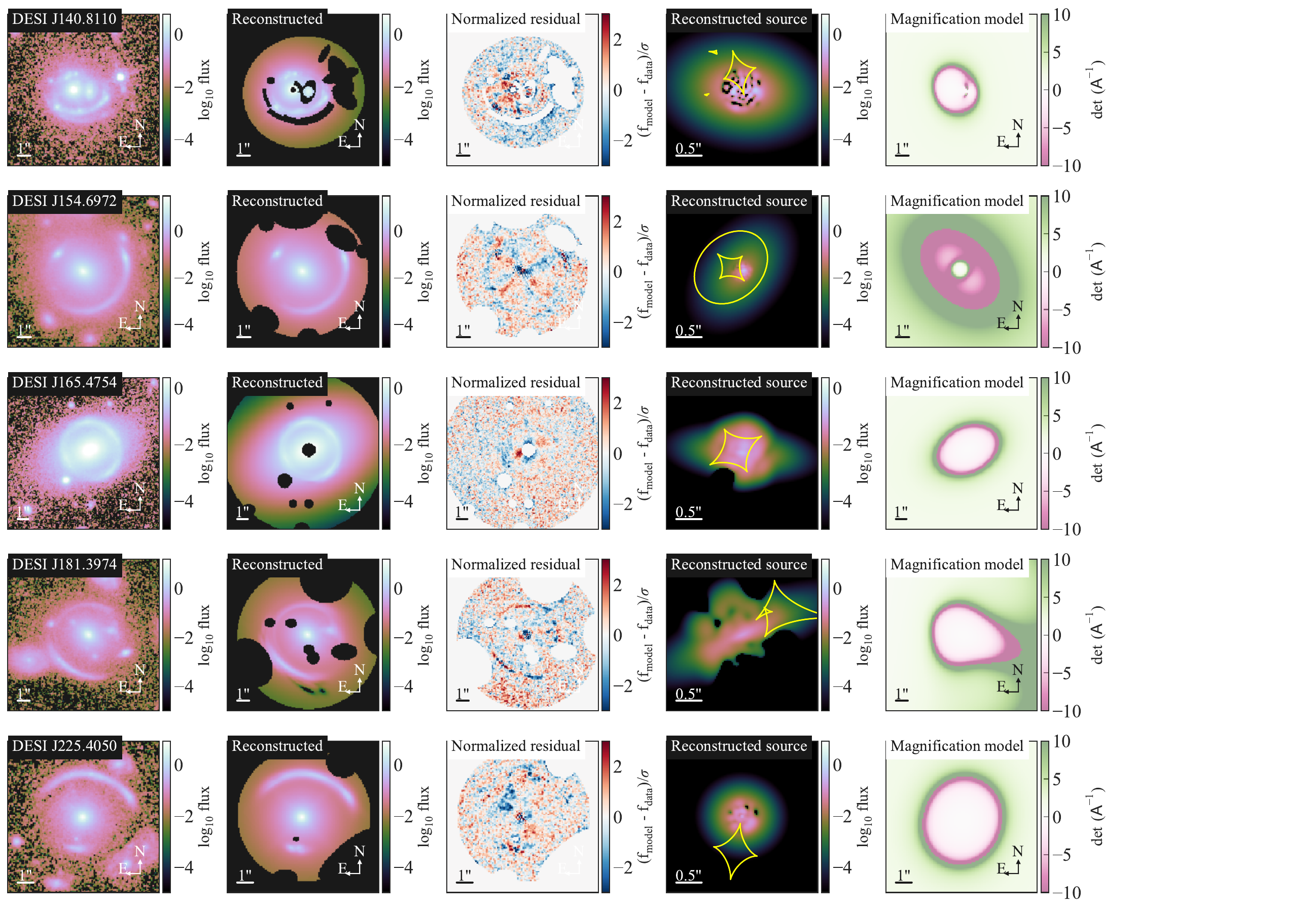}
	\caption{\label{fig:lens_models_1}
    Illustration of lens models for the second five out of 15 systems in our sample (continued from Fig. \ref{fig:lens_models_0}).
	}
\end{figure*}

\begin{figure*}[h!]
	\centering
	\includegraphics[width=1.3\textwidth]{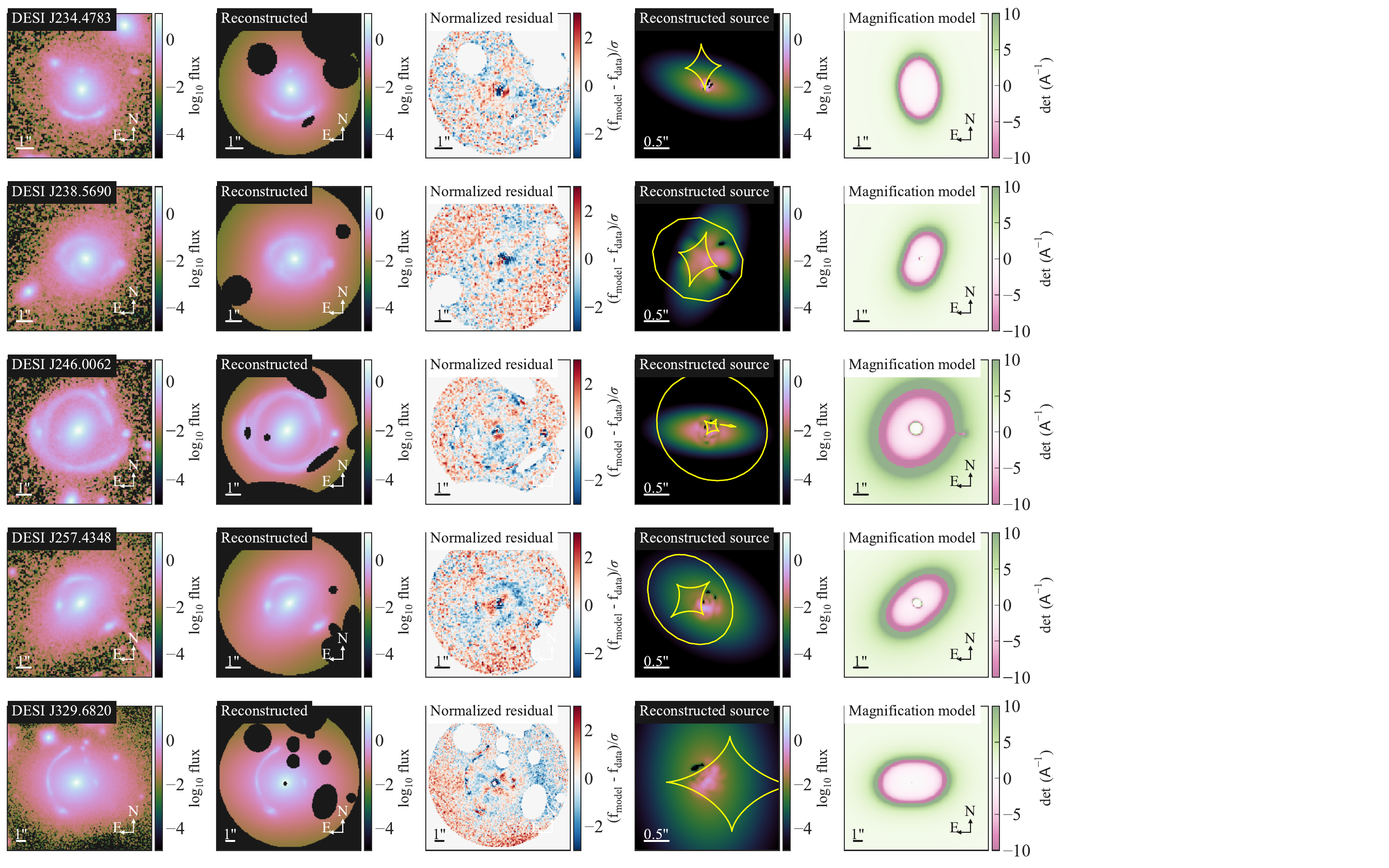}
	\caption{\label{fig:lens_models_2}
	  Illustration of lens models for the last five out of 15 systems in our sample (continued from Figures \ref{fig:lens_models_0} and \ref{fig:lens_models_1}).
	}
\end{figure*}

\finaledits{In Figures \ref{fig:lens_models_1} and \ref{fig:lens_models_2}, we show the image cutouts, reconstructed images, residual maps, reconstructed flux distribution of the source galaxies, and the magnification models for the remaining models that were not included in Fig. \ref{fig:lens_models_0}.} 


\end{appendix}

\end{document}